\documentclass[%
 reprint,
superscriptaddress,
 amsmath,amssymb,
 aps,
floatfix,
]{revtex4-2}

\newcommand{\narrowbaselines}{\relax}
\usepackage{tex_elements/tysty}
\usepackage{graphicx}% Include figure files
\usepackage{dcolumn}% Align table columns on decimal point
\usepackage{bm}% bold math
\usepackage{algorithmic}
\usepackage{algorithm}
\DeclareMathOperator{\origin}{or}
\DeclareMathOperator{\terminus}{ter}

\PassOptionsToPackage{hyphens}{url}
\renewcommand{\red}{\relax}

\begin{document}

\preprint{APS/123-QED}

\title{Quantum spatial best-arm identification on a complete bipartite graph}% Force line breaks with 

\author{Tomoki Yamagami}
\email{tyamagami@mail.saitama-u.ac.jp}
\affiliation{Department of Information and Computer Sciences, Saitama University,\\
255 Shimo-Okubo, Sakura-ku, Saitama City, Saitama 338--8570, Japan.}%}
\affiliation{Department of Information Physics and Computing, The University of Tokyo,\\
7--3--1 Hongo, Bunkyo-ku, Tokyo 113--8656, Japan.}
\author{Etsuo Segawa}
\affiliation{Graduate School of Environment and Information Sciences,
Yokohama National University,\\
79--1 Tokiwadai, Hodogaya-ku, Yokohama City, Kanagawa 240--8501, Japan.}
\author{Takatomo Mihana}
\affiliation{Department of Information Physics and Computing, The University of Tokyo,\\
7--3--1 Hongo, Bunkyo-ku, Tokyo 113--8656, Japan.}
\author{Andr\'e R\"ohm}
\affiliation{Department of Information Physics and Computing, The University of Tokyo,\\
7--3--1 Hongo, Bunkyo-ku, Tokyo 113--8656, Japan.}
\author{\\Atsushi Uchida}
\email{auchida@mail.saitama-u.ac.jp}
\affiliation{Department of Information and Computer Sciences, Saitama University,\\
255 Shimo-Okubo, Sakura-ku, Saitama City, Saitama 338--8570, Japan.}%}
\author{Ryoichi Horisaki}
\affiliation{Department of Information Physics and Computing, The University of Tokyo,\\
7--3--1 Hongo, Bunkyo-ku, Tokyo 113--8656, Japan.}

\begin{abstract}
Quantum reinforcement learning has emerged as a framework combining quantum computation with sequential decision-making, and applications to the multi-armed bandit (MAB) problem have been reported.
The graph bandit problem extends the MAB setting by introducing spatial constraints, where the accessibility of arms is restricted by graph connectivity,
yet quantum approaches to this setting remain limited.
In this paper, we formulate best-arm identification in graph bandits and propose a quantum algorithmic framework, termed \textit{Quantum Spatial Best-Arm Identification} (QSBAI), which is applicable to general graph structures.
This framework uses quantum walks to encode superpositions over graph-constrained actions,
thereby extending amplitude amplification and generalizing the quantum BAI algorithm via Szegedy's walk framework.
We focus our theoretical analysis on complete and bipartite graphs, deriving the maximal success probability of identifying the best arm and the time step at which it is achieved.
Our results clarify how quantum-walk-based search can be adapted to structurally constrained decision problems
and provide a foundation for quantum best-arm identification in graph-structured environments.
\end{abstract}

\maketitle

%\tableofcontents

\section{\label{intro}Introduction}%\setcounter{equation}{0}
\textit{Reinforcement learning}~\cite{sutton2018reinforcement} has become an increasingly crucial paradigm in machine learning.
Reinforcement learning is a subfield of machine learning focused on solving sequential decision-making problems, where a learner, or an \textit{agent}, interacts with an environment to learn an optimal policy for selecting actions. 
The agent explores possible actions, receives feedback in the form of rewards or penalties, and gradually refines its strategy.
Reinforcement learning has become a fundamental framework for adaptive decision-making in diverse fields, including robotics \cite{hwangbo2019learning}, finance \cite{kim2019optimizing}, and automation \cite{zhao2020deep}.
Recent advancements in reinforcement learning algorithms have significantly broadened its applicability, driving both research interest and practical adoption.

\begin{figure}
    \centering
    \includegraphics[width=68.5mm]{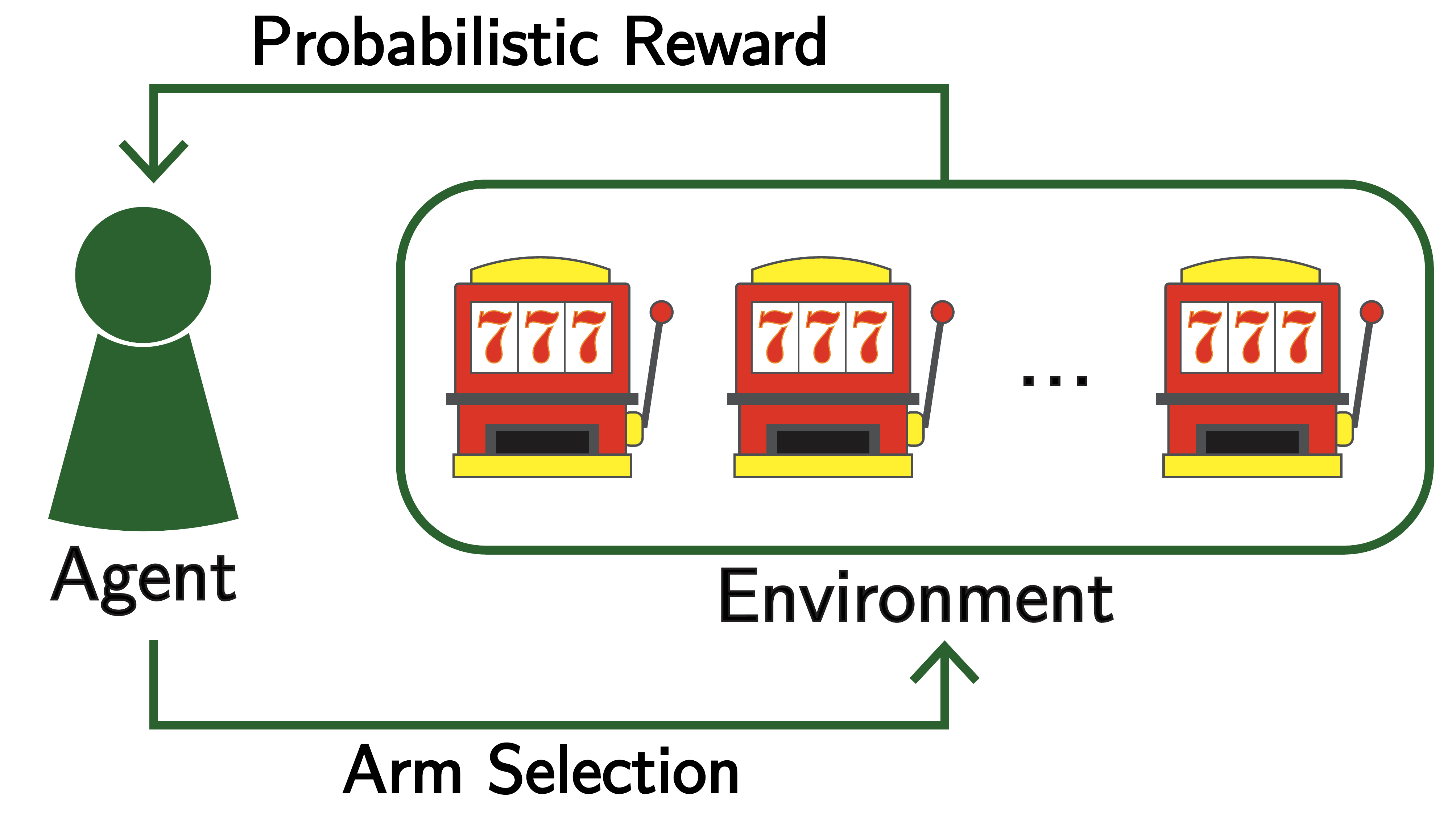}
    \caption{Conceptual illustration of a multi-armed bandit (MAB) problem.
    The MAB framework consists of interactions between an agent and an environment containing multiple arms (slot machines).
    The agent selects an arm, which then generates a reward probabilistically.
    Based on the observed reward, the agent updates its strategy and selects an arm again.}\label{intro:fig:mab}
\end{figure}

A \textit{Multi-Armed Bandit} (MAB) problem~\cite{robbins1952some,slivkins2019introduction} is a fundamental framework of reinforcement learning.
In MAB problems, it is assumed that the environment has multiple selections or \textit{arms}, and a single decision is represented by selecting an arm, after which a reward is generated depending on which arm is selected, as shown in Fig.~\ref{intro:fig:mab}.
This reward is determined by a distribution that each arm has independently; however, the agent does not possess any information about the distributions of the arms.
Therefore, one needs to learn which arm is expected to produce the larger reward during multiple selections, or which is the \textit{best arm}, while practically receiving rewards.

In many cases, MAB problems consider the scenario where the agent seeks to minimize regret, defined as the difference between the expected total reward under optimal selections and the actual total reward.
On the other hand, there is another scenario: one aims at identifying the best arm as quickly as possible—referred to as \textit{Best-Arm Identification} (BAI) problems \cite{bubeck2009pure,audibert2010best}.
In regret minimization, it is necessary to allocate resources to the empirically best arm in order to obtain rewards stably, while still exploring other arms, where the dilemma called the \textit{exploration–exploitation trade-off} \cite{march1991exploration} is inherent.
By contrast, in BAI problems, the agent focuses solely on judging the best arm with as little time or budget as possible; that is, only exploration is involved.

Variants of MAB problems arise when structural constraints are introduced into the environment. 
One important example is the \textit{graph bandit setting}, where arms are represented by the nodes of a graph, and choices are restricted by the graph's connectivity~\cite{johansson2022graph,zhang2023multi}. 
That is, it is assumed that the agent makes choices while traveling on the graph, and the next choice is limited to the neighborhood of the currently selected arm.
In the conventional MAB problems, it is treated as a matter of course that the agent is free to access any arm in the environment.
In real problems, however, the effective constraint of the current selection on the next should be considered, as well as the expectation of a reward of the selection.
For example, MAB problems are applied to wireless communication such as channel selection and beam alignment~\cite{li2020multi,takeuchi2020dynamic,zhang2020beam}.
In these applications, the graph structure can represent either abstract relations, such as interference among channels, or concrete spatial relations, such as the adjacency of cells or beam directions.
Accordingly, the assumption that the agent moves on the graph captures realistic limitations: a device scanning channels must proceed sequentially across adjacent frequencies, or a mobile terminal can only hand over to neighboring cells.
Another example is presented in terms of asset management~\cite{johansson2022graph};
here, each node of the graph represents a portfolio consisting of long, short, or neutral positions in a set of assets, and the edges correspond to single adjustments of these positions.
Thus, the agent's movement on the graph reflects realistic portfolio management, where changes are incremental rather than arbitrary, and the rewards correspond to the stochastic returns of the portfolio.
Graph bandit settings are often related to contextual bandit formulations~\cite{auer2002using}, where the context is provided by adjacency relations among arms, and are widely applicable to decision-making tasks influenced by external information.
\red{Under such locality constraints, BAI is viewed as comparing arms by rewards while exploring a structured action space under restricted accessibility.}

One important direction of developing MAB problems includes constructing quantum solutions.
This is related to the fact that one emerging frontier in reinforcement learning is \textit{quantum reinforcement learning}, a part of quantum machine learning \cite{briegel2012projective,wittek2014quantum,aaronson2015read,dunjko2016quantum,lamata2017basic,biamonte2017quantum}, which leverages quantum computing principles to enhance decision-making efficiency and scalability \cite{dong2008quantum,paparo2014quantum,li2020quantum}. 
Quantum computing itself~\cite{nielsen2010quantum,bennett2000quantum} is a rapidly developing paradigm, originating with Feynman's proposal of quantum simulation \cite{feynman1982simulating} and Deutsch's principle of quantum parallelism \cite{deutsch1985quantum}. 
Seminal algorithms such as Shor's factorization algorithm \cite{shor1994algorithms} and Grover's quantum search algorithm \cite{grover1996fast} demonstrated exponential or quadratic speedups over classical methods, highlighting the transformative potential of quantum computation. 
Building on these foundations, quantum reinforcement learning has been proposed as a way to achieve more efficient decision-making by exploiting quantum parallelism and entanglement \cite{dong2008quantum,paparo2014quantum,li2020quantum}, although practical applications remain limited by current hardware constraints.

Several quantum solutions for BAI problems or related problems have been proposed as kinds of quantum reinforcement learning algorithms.
BAI problems are well-suited for quantum search algorithms, including Grover's algorithm \cite{grover1996fast} and its generalization, referred to as quantum amplitude amplification \cite{brassard2002quantum}, since BAI problems do not involve exploitation and concentrate only on exploration.
Indeed, Casal\'e et al. \cite{casale2020quantum} proposed an algorithm named \textit{Quantum BAI} (QBAI) as an application of quantum amplitude amplification and theoretically showed that QBAI obtains optimal solutions quadratically faster than a classical policy.
Wang et al. \cite{wang2021quantum_A} also presented a quantum-amplitude-amplification-based BAI algorithm with a different approach, showing that the best arm can be found with a fixed confidence.
Moreover, Cho et al. \cite{cho2023quantum} proposed a decision-making algorithm in the framework of adversarial MAB problems \cite{auer2002nonstochastic}, which is also exploration-based with quantum amplitude amplification.

\begin{figure}
    \centering
    \includegraphics[width=68.5mm]{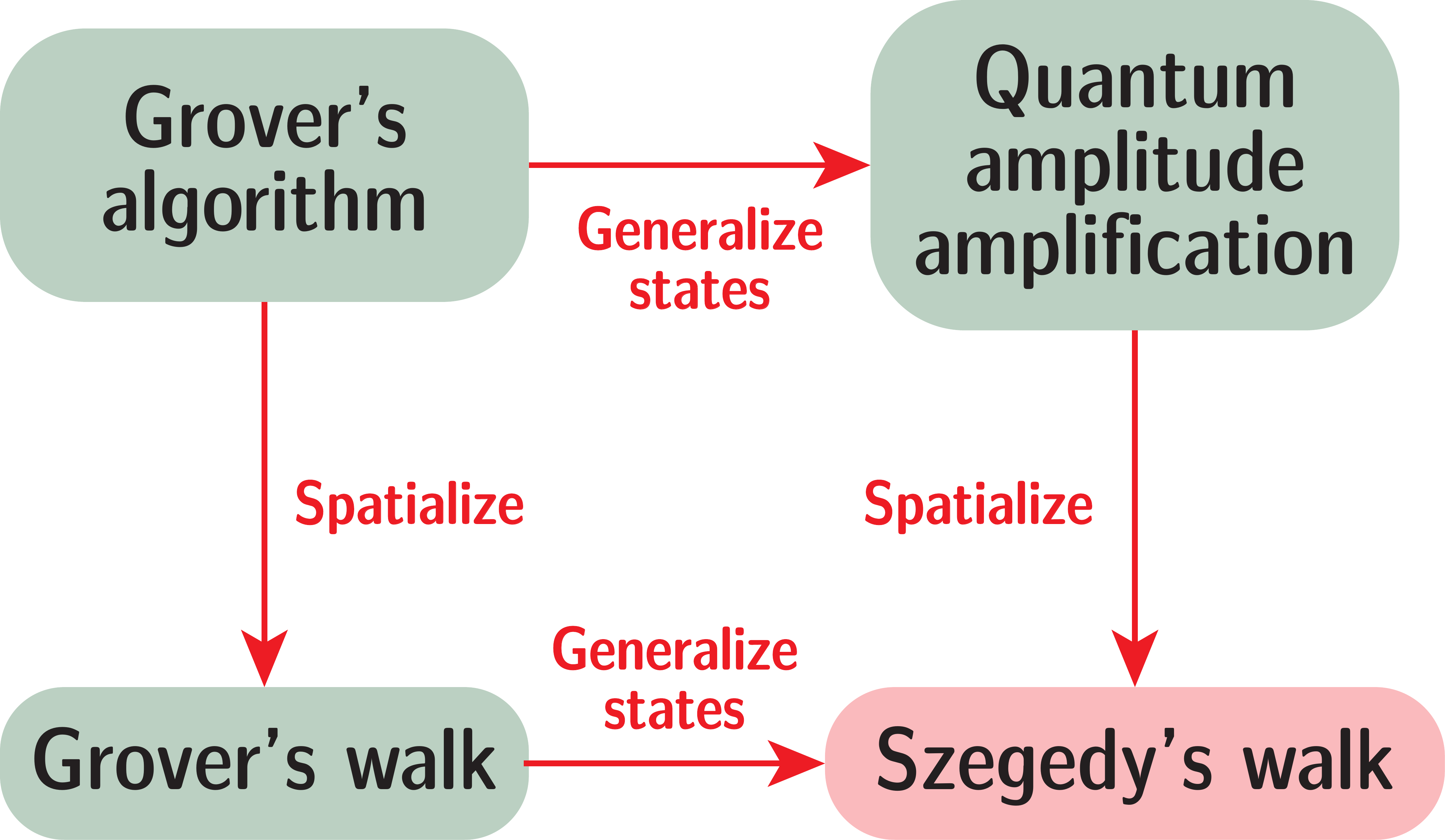}
    \caption{Positioning of Szegedy's walk.
    Grover's search algorithm has two direction of generalization: regarding the initial state and spatial constraints.
    The former and latter are called quantum amplitude amplification and Grover's walk, respectively.
    Szegedy's walk is positioned on the intersection of these two ways of generalizing Grover's walk.}\label{intro:fig:szegedy}
\end{figure}

\red{We propose a quantum algorithm for BAI problems in a graph-bandit setting, which we call \textit{Quantum Spatial Best-Arm Identification} (QSBAI)~\cite{yamagami2023quantum}.
The essential extension from the conventional QBAI in the previous work~\cite{casale2020quantum} is that arm accessibility is no longer unrestricted, and
the next decision is constrained by the graph structure.
Our goal is therefore to formulate BAI under such restricted accessibility and to construct a quantum search procedure adapted to this setting.
To describe quantum superposition over graph-structured arm selections, we employ quantum spatial search by \textit{Quantum Walks} (QWs) on graphs.}
QWs are the quantum version of random walks, which considers quantum superposition about state transition~\cite{konno2008quantum,portugal2013quantum}. 
QWs are useful for constructing and analyzing quantum algorithms \cite{kempe2003quantum,konno2008quantum,venegas2012quantum,portugal2013quantum}; for instance, it has been clarified that Grover's algorithm can be understood as a specific case of spatial search \cite{portugal2013quantum} by incorporating the operator corresponding to the quantum oracle into a class of QWs called Grover's walk.
Moreover, quantum amplitude amplification, which generalizes the initial state and the operator referring the initial state in Grover's algorithm, is also covered by a Markov-process-based QW called Szegedy's walk \cite{szegedy2004quantum}.
Grover's walk can be captured as a special case of Szegedy's walk; in that sense, Szegedy's walk can be positioned as a model which generalizes Grover's algorithm in terms of both the initial state and spatial structure, also see Fig.~\ref{intro:fig:szegedy}.
This paper attempts to construct QSBAI by generalizing quantum amplitude amplification that appears in QBAI with Szegedy's walk.
Also, we clarify that our proposed algorithm is a natural extension of the conventional QBAI algorithm without spatial structure \cite{casale2020quantum} through presenting an example of QSBAI on complete graphs.

\red{We investigate QSBAI on complete bipartite graphs and derive analytical estimates of its performance. Complete bipartite graphs provide a useful model of structured accessibility, in which the action space is partitioned into two classes and admissible transitions are governed by this partition. Such graph structures are relevant to engineering decision problems, in which physical or operational constraints restrict how the agent moves through possible actions under locality constraints~\cite{kim2022bipartite,dai2024survey}. Complete bipartite graphs therefore provide a setting for investigating best-arm identification under graph-constrained exploration. They also offer an analytically tractable setting, in which the effect of accessibility constraints on QSBAI can be examined explicitly.}

\if0 %%%commenting out%%%
\begin{figure}
\centering
\begin{minipage}[b]{68.5mm}
    \centering\includegraphics[width=68.5mm]{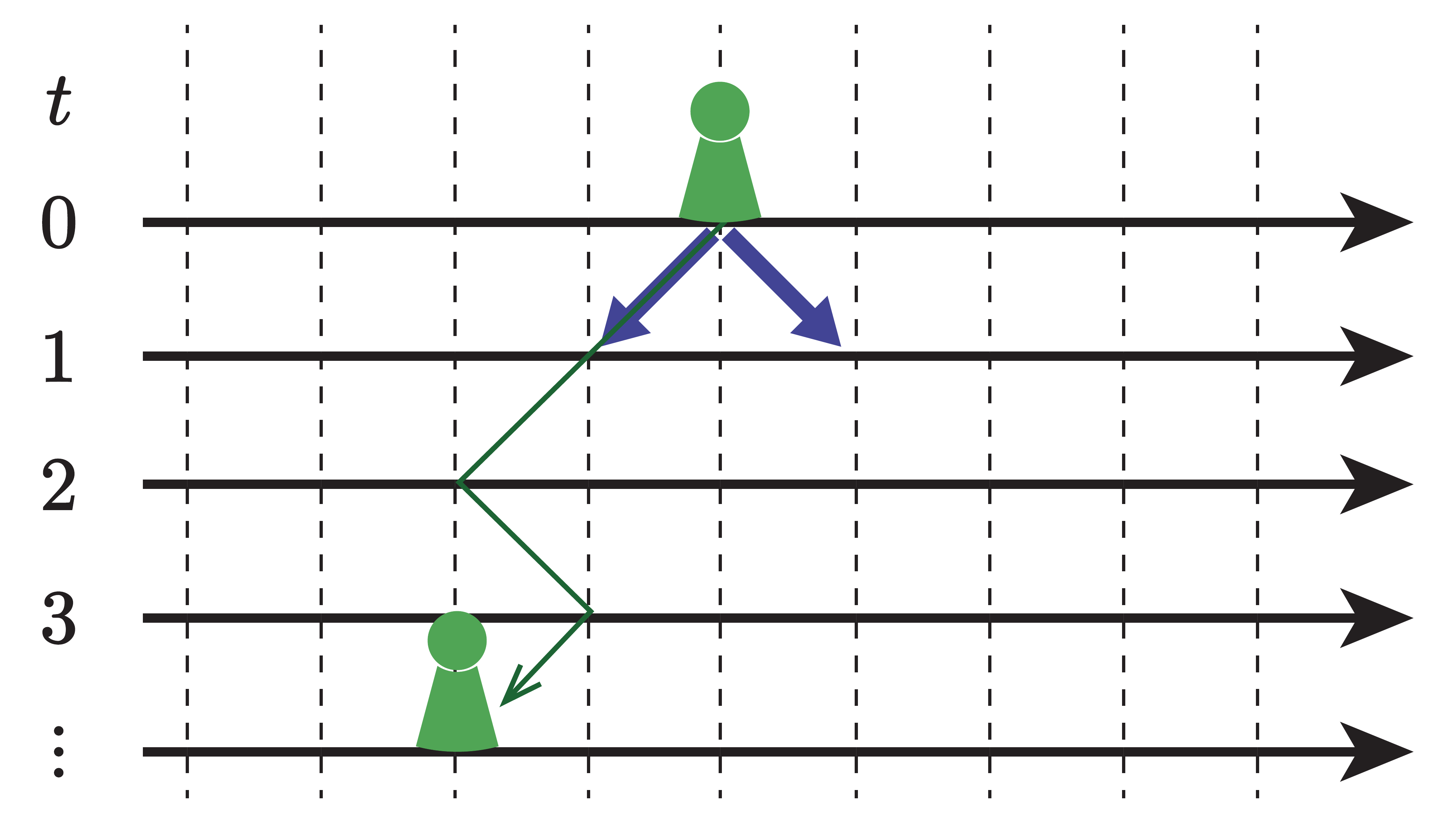}
    \text{(a)\ Random walks.}\vrule height0pt width0pt depth8pt
\end{minipage}\\
\begin{minipage}[b]{68.5mm}
    \centering\includegraphics[width=68.5mm]{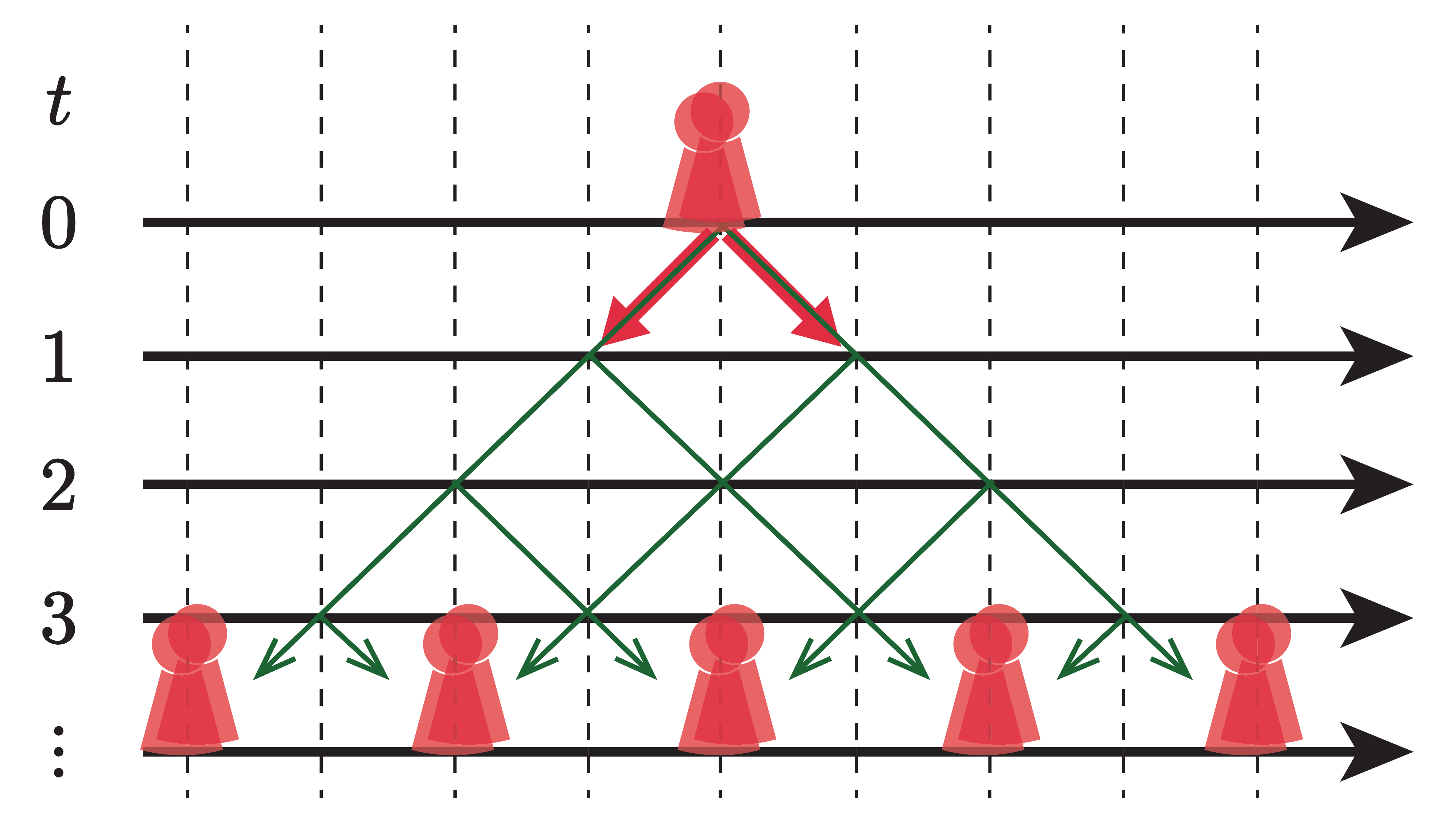}
    \text{(b)\ Quantum walks.}
\end{minipage}
\caption{Conceptual illustration of the difference between random walks and quantum walks on a one-dimensional lattice.
    Whereas random walks are characterized by repeated stochastic choices of left or right directions, quantum walks are described by quantum superpositions of possible path trajectories.
    A quantum walker can occupy multiple positions simultaneously until measurement.}\label{intro:fig:qw}
\end{figure}
\fi %%%%%%%%%%%%%%%%%%%%%%

Our study lies at the intersection of quantum walks (QWs), the quantum best-arm identification (QBAI) algorithm, and the graph-structured bandit framework.
In terms of the fusion of QWs and MAB algorithms, previous work has proposed a QW-based policy that combines the coexisting properties of linear spreading and localization with the two aspects that must be balanced, exploration and exploitation \cite{yamagami2023bandit}.
However, our approach is clearly distinct from the previous one in both the problem setting and the utilization of QWs during the decision-making process.
The earlier study focuses on maximizing cumulative rewards over a finite number of decision rounds, where the exploration–exploitation trade-off is central.
By contrast, our study targets the identification of the best arm, independent of this trade-off.
Another key difference is the assumption that arms are embedded in a graph structure, which further distinguishes our framework.
Regarding the use of QWs, the previous study applies a QW---comprising time evolution and measurement---for each decision, adjusting the QW setting at every step.
On the other hand, our model employs a single sequence of QW operations throughout the algorithm, culminating in just one measurement at the end.

This paper is organized as follows.
Section \ref{qwss} reviews QWs on arbitrary graphs and the spatial search algorithm therein.
Based on the notations introduced there, we propose the QSBAI algorithm in Sec.~\ref{qsbai}.
Section \ref{exampleC} presents QSBAI on complete graphs with self-loops.
Section \ref{exampleBP} gives QSBAI on complete bipartite graphs as one case of the simplest examples with spatial structure.
Section \ref{discussion} discusses open problems and future work for further studies.
Section \ref{conclusion} concludes this paper.

%%%%%%%%%%%%%%%
%%%%%%%%%%%%%%%

\section{\label{qwss}Quantum walk on graphs and spatial search}%\setcounter{equation}{0}
This section reviews QWs on graphs $G$ and search algorithms based on QWs. 
Further details on these topics can be found in Ref.~\cite{portugal2013quantum}.

\subsection{\label{qwss:sub:qw}Quantum walks on graphs}
Here we give a standard form of quantum walks on graphs.
Let $G = (V,\,A)$ be a symmetric digraph, where $V$ is the set of vertices and $A \subset V\times V$ is the set of directed edges, or \textit{arcs}, satisfying the condition that for any $a = (u,\,v)\in A$, the inverse arc $a^{-1} := (v,\,u)$ is also contained in $A$.
For $a = (u,\,v)\in A$, $u$ (resp. $v$) is called the origin (resp. terminus) of the arc $a$, denoted by $\origin(a)$ (resp. $\terminus(a)$).

The system of QWs on a symmetric digraph $G$ is defined on the Hilbert space $\mathcal{H}_{A}$ spanned by the arcs of $G$:
\begin{equation}
    \mathcal{H}_A = \spn \{ \ket{a} \,|\,a\in A\}.
\end{equation}
We define $\{\ket{a}\,|\,a\in A\}$ to satisfy
\begin{equation}
    \braket{a|b} = \delta_{ab} := \begin{cases} 1 & (a=b), \\ 0 & (a\neq b). \end{cases}
\end{equation}
Thus, $\{\ket{a}\,|\,a\in A\}$ forms an orthonormal system, and $\mathcal{H}_A$ is equivalent to
\begin{equation}
    \ell^2(A) = \left\{ f: A\to \mathbb{C}\,\middle|\, \sum_{a\in A}|f(a)|^2 < \infty \right\}.
\end{equation}
A state of a QW is defined as a unit vector in $\mathcal{H}_A$. For any state $\ket{\Psi}$ of a QW on $G$, there exists a collection $(\alpha_a\,|\,a\in A)$ of complex numbers such that
\begin{equation}
    \ket{\Psi} = \sum_{a\in A} \alpha_a \ket{a} \in \mathcal{H}_A
\end{equation}
with
\begin{equation}\label{qwss:eq:whole}
    \sum_{a\in A}|\alpha_a|^2 = 1.
\end{equation}
Here, $\alpha_a$ is called the \textit{probability amplitude} of arc $a$, and it satisfies $\alpha_a = \braket{a|\Psi}$.

States of QWs mathematically represent particles that exist non-locally on the graph $G$. 
The position of a particle is probabilistically determined through measurement.
The probability that the particle is found at vertex $v\in V$ under the state $\ket{\Psi}$ is
\begin{equation}
    \mu(v) = \sum_{a\in \mathcal{T}(v)}\left|\braket{a|\Psi}\right|^2 = \sum_{a\in \mathcal{T}(v)}\left|\alpha_a\right|^2,
\end{equation}
where $\mathcal{T}(v):= \{a\in A\,|\,\terminus(a) = v\}$ is the set of arcs terminating at $v$.
By Eq.~\eqref{qwss:eq:whole}, we have $\sum_{v\in V}\mu(v) = 1$, confirming that this is a probability distribution.

Next, we describe the variation of the state of a QW, referred to as its \textit{time evolution}.
Let $t\in \{0\}\cup \mathbb{N} =: \mathbb{N}_0$ denote the number of steps in the time evolution. 
We denote the state of the QW at time step $t\in \mathbb{N}_0$ by $\ket{\Psi_t}$ and assume that the time evolution of the state is governed by a unitary operator $U$ on $\mathcal{H}_A$ as
\begin{equation}\label{qwss:eq:tev}
	\ket{\Psi_{t+1}} = U\ket{\Psi_t}
\end{equation}
for any $t$. 
The time evolution operator $U$ is defined as
\begin{equation}\label{qwss:eq:U}
	U = \sum_{a\in A}\sum_{b\in\mathcal{T}(\terminus(a))}\gamma_{ab}\ketbra{b^{-1}}{a},
\end{equation}
where $\gamma_{ab} = (U)_{b^{-1}a}\in \mathbb{C}$, and for any $v\in V$, the $|\mathcal{T}(v)|$-dimensional matrix $[\gamma_{ab}]_{a, b\in\mathcal{T}(v)}$ is unitary.
The sequence $(\Psi_t\,|\,t\in\mathbb{N}_0)$ of QWs on graph $G$ is determined by the initial state, set to a unit vector $\ket{\Phi} \in\mathcal{H}_A$, and the time evolution operator $U$ defined by Eq.~\eqref{qwss:eq:U}.

\subsection{\label{qwss:sub:ss}Spatial search via QWs}
The spatial search on a graph $G$ aims to find specific vertices, referred to as marked vertices, embedded in $G$.
We introduce a subset $W\subset V$ as the set of marked vertices and define a map $f: V\to \{0,\, 1\}$ as
\begin{equation}
    f(v) := \mathbf{1}_W(v) = \begin{cases} 1 & (v\in W), \\0 & (v\in V\setminus W), \end{cases}
\end{equation}
where $\mathbf{1}_S$ is the indicator function of set $S$.
This map $f$ is called the classical oracle, which tells whether the input vertex is marked.
Using this notation, the aim of spatial search can be expressed as maximizing the probability of locating particles on marked vertices, achieved via quantum amplitude amplification through the time evolution of QW states. 

In spatial search algorithms, we consider the coin operator composed of two unitary operators $U_0$ and $R_f$ as
\begin{equation}\label{qwss:eq:Csearch}
    U = U_0R_f.
\end{equation}
Here, $U_0$ is a ``standard'' unitary operator~\cite{portugal2013quantum}; in the original spatial search, for example, it is defined as Grover’s matrix \cite{grover1996fast}, which sets
\begin{equation}
    U_0 = \sum_{a\in A}\sum_{b\in \mathcal{T}(\terminus(a))}\gamma_{ab}'\ketbra{b^{-1}}{a}
\end{equation}
with
\begin{equation}
    \gamma_{ab}' = \frac{2}{\deg(\terminus(b))} -\delta_{ab}.
\end{equation}
In QSBAI, presented in Sec.~\ref{qsbai}, $U_0$ is defined in Szegedy’s manner \cite{szegedy2004quantum}, based on a Markov process derived from classical decision-making.

The other component, $R_f$, is often called the quantum oracle and is defined as
\begin{equation}
    \begin{split}
    R_f &= \sum_{a\in A} (-1)^{f(\terminus(a))}\ketbra{a}{a} \\
    &= \sum_{u\in V}\sum_{v\in\mathcal{N}_G(u)} (-1)^{f(v)}\ketbra{(u,\,v)}{(u,\,v)}.
    \end{split}
\end{equation}
It incorporates the information about marked vertices by inverting the sign of states corresponding to marked vertices~\cite{portugal2013quantum}.
In QW-perspectives, inserting $R_f$ corresponds to providing specific vertices with ``defects," which can result in concentrating measurement probability on the vertices when one makes the measurement at a proper time.

After $t$ steps of time evolution with the unitary operator $U$ defined in Eq.~\eqref{qwss:eq:U}, where the coin operator is given by Eq.~\eqref{qwss:eq:Csearch}, a measurement is performed. Then, denoting the QW state at time step $t$ by $\ket{\Psi_t}$, the probability of finding one of the marked vertices in $G$ is
\begin{equation}
    \begin{split}
    P_t^{(*)} = \sum_{w\in W}\mu(w) &= \sum_{w\in W}\sum_{a\in \mathcal{T}(w)}\left|\braket{a|\Psi_t} \right|^2\\
    &= \sum_{w\in W}\sum_{a\in \mathcal{T}(w)}\left|\braket{a|U^t|\Psi_0} \right|^2.
    \end{split}\label{qwss:eq:measurement}
\end{equation}
This is the objective in the original spatial search problem; however, our optimization target via quantum amplitude amplification differs, as described in Sec.~\ref{qsbai}.

\subsection{\label{qwss:sub:szegedy}Szegedy's walk}
Szegedy’s walk is a class of QWs derived from a method of quantizing Markov processes or classical random walks, introduced by Szegedy \cite{szegedy2004quantum,portugal2013quantum,higuchi2017periodicity}. 
More precisely, a Szegedy walk on graph $G$ is constructed from a transition probability $p:A\to [0,\,1]$ on $G$ and a \textit{reversible distribution} $\pi: V\to [0,\,1]$ associated with $p$. 
If $a = (u,\,v)$, then $p(a) = p(u,\,v)$ represents the transition probability from vertex $u$ to $v$, and a reversible distribution $\pi$ is a probability distribution satisfying
\begin{equation}\label{qsww:eq:dbc1}
	p(u,\,v) \pi(u) = p(v,\,u) \pi(v)
\end{equation}
for any pair $u,\ v\in V$. 
That is, for any arc $a\in A$, $\pi$ satisfies
\begin{equation}\label{qsww:eq:dbc2}
	p(a)\pi(\origin(a)) = p(a^{-1})\pi(\terminus(a)).
\end{equation}
Equations~\eqref{qsww:eq:dbc1} and \eqref{qsww:eq:dbc2} are referred to as the detailed balance condition between vertices $u$ and $v$. 
Moreover, the function $m:A\to [0,\,1]$ defined by $m(u,\,v) := p(u,\,v)\pi(u)$ represents the flow of probability from vertex $u$ to $v$; the detailed balance condition indicates that the inflow and outflow between $u$ and $v$ are equal.

As noted above, QWs are constructed by defining the initial state $\ket{\Psi_0} = \ket{\Phi}$ and the time-evolution operator $U$. 
First, Szegedy’s walk sets the initial state, weighted by the probability flow $m(a)$, to
\begin{align}
	\ket{\Phi} &= \sum_{a\in A} \sqrt{m(a)}\ket{a}.
\end{align}
Furthermore, the time-evolution operator $U$ is defined by setting the parameters $\gamma_{ab}$ in terms of the transition probabilities associated with the inverse arcs of $a$ and $b$ as
\begin{equation}\label{qsww:eq:gamma}
	\gamma_{ab} = \begin{cases} 
 2p(a^{-1}) -1 & (a=b), \\[4pt]
 2\sqrt{p(a^{-1})p(b^{-1})} & (\text{otherwise}). 
 \end{cases}
\end{equation}
Recall that $a$ and $b$ are assumed to share a vertex as their termini, by the definition of $U$ in Eq.~\eqref{qwss:eq:U}. 
It should be emphasized that the normalization of $\ket{\Phi}$ and the unitarity of $U$ are ensured by the requirements that $p$ is a transition probability and $\pi$ is a probability distribution:
\begin{align}
	\sum_{a:\origin(a) = u}p(a) &= \sum_{v\in V}p(u,\,v) = 1\quad\text{for any $u\in V$},  \label{qwss:eq:unity_pa}\\
	\sum_{v\in V} \pi(v) &= 1. \label{qwss:eq:unity_pi}
\end{align}

\red{This method is not introduced independently of the decision problem in QSBAI. The Szegedy's walk is constructed from a classical state-transition rule induced by graph-constrained arm selection and stochastic environment states, and thus provides a framework for connecting spatially constrained decision making to quantum amplitude amplification.}

%%%%%%%%%%%%%%%
%%%%%%%%%%%%%%%

\section{\label{qsbai}Quantum spatial best-arm identification}%\setcounter{equation}{0}
In this section, we formalize best-arm identification under spatial constraints and then present the corresponding quantum policy, which we call \textit{Quantum Spatial Best-Arm Identification (QSBAI)}.

\subsection{\label{qsbai:sub:problem}Problem setting}
Let $G=(V,\,A)$ be a graph representing the environment.
Here, $V$ is the set of vertices in $G$, and an arm is placed on each vertex. 
In the following, we identify the arm located at vertex $v\in V$ with $v$ itself and refer to it as arm $v$.

We begin by describing a single decision-making scenario in the classical context, which serves as the basis of our quantum strategy. 
Assume that an agent is located at a vertex $v_0$ in graph $G$. 
The agent selects one of the arms positioned at the vertices adjacent to $v_0$.
Note that this selection rule differs from that in MAB problems without spatial constraints: in our setting, the agent cannot select arms that are not connected to the current vertex, whereas in conventional MAB problems, the agent may select from all arms according to its policy.
After selecting and playing an arm, the environment probabilistically generates a reward according to the probability distribution associated with the chosen arm.
We assume that the reward distribution for each arm $v$ follows a Bernoulli distribution with parameter $q_v$.
That is, when arm $v$ is selected, the agent receives a unit reward (i.e., the arm \textit{wins}) with probability $q_v$.
We refer to $q_v$ as the \textit{winning probability} of arm $v$.
This winning probability $q_v$ is fixed for each $v\in V$, but the agent cannot directly observe it and must instead learn it empirically through repeated selections.

In standard MAB problems, the winning probabilities of arms are directly specified and form the basis for decision-making schemes. 
Such settings differ fundamentally from conventional quantum search algorithms. 
In quantum search, each vertex of a graph is deterministically marked or unmarked; in other words, the outcome of selecting a vertex is always consistent \cite{grover1996fast}. 
In our setting, however, such consistency is absent due to the probabilistic nature of each arm’s winning probability.

To bridge this gap, we adopt the notion of \textit{environment states} introduced in Ref.~\cite{casale2020quantum}. 
The environment state is a random variable, not directly observable by the agent, that follows the probability distribution $\eta_v$ associated with the selected arm $v\in V$. 
This concept formalizes the probabilistic determination of rewards.
Let $\Sigma$ be the set of possible environment states. 
Whether arm $v$ wins or loses is determined by the environment state $\sigma$ that arises after selecting $v$. 
Formally, the reward law is determined by pairs $(v,\,\sigma)\in V\times \Sigma$. 
We denote the subset of winning pairs as $W\subset V\times \Sigma$ and define the map $f:V\times \Sigma \to \{0,\,1\}$ as
\begin{equation}\label{qsbai:eq:f}
	f(v,\,\sigma) = \mathbf{1}_W(v,\,\sigma) = 
 \begin{cases} 
 1 & \text{if } (v,\,\sigma)\in W, \\[4pt]
 0 & \text{if } (v,\,\sigma)\notin W. 
 \end{cases}
\end{equation}
Here, $f(v,\,\sigma) = 1$ (resp. $0$) means that the selected arm $v$ wins (resp. loses) under environment state $\sigma$.
Under this setting, the winning probability of arm $v$ can be expressed as
\begin{equation}
	q_v = \sum_{\sigma\in \Sigma_v^*}\eta_v(\sigma),
\end{equation}
where $\Sigma_v^*$ denotes the set of environment states under which arm $v$ wins:
\begin{equation}
	\Sigma_v^* := \{\sigma\in\Sigma\,|\,(v,\,\sigma)\in W\}.
\end{equation}

One round of decision-making can now be described as follows. 
First, the agent selects an arm $v\in V$ adjacent to the current arm $v_0$. 
Next, the environment state $\sigma\in \Sigma$ is sampled according to the distribution $\eta_v$. 
Finally, the reward is determined as $r = f(v,\,\sigma)$. 
The goal of the BAI algorithm is to maximize the probability of selecting the \textit{best arm} $v^\star\in V$ with as few rounds as possible. 
The best arm $v^\star$ is defined as the arm with the highest winning probability:
\begin{equation}\label{qsbai:qw:bestarm}
    v^\star = \argmax_{v\in V} q_v.
\end{equation}
We assume that the best arm is unique for simplicity.

\subsection{\label{qsbai:sub:method}Proposed Method}
In this subsection, we propose Quantum Spatial Best-Arm Identification (QSBAI), the BAI algorithm under the setting described in Sec.~\ref{qsbai:sub:problem}.

QSBAI is constructed in two stages.
First, we describe the classical state-transition structure induced by graph-constrained arm selection and stochastic environment states.
Then, we quantize this transition structure through Szegedy's walk to obtain a quantum policy for best-arm identification under spatial constraints.
The pseudocode of QSBAI is presented in Algorithm~\ref{qsbai:alg}.
In the following, we provide a detailed discussion leading to this algorithm.

\begin{figure}[ht]
\begin{algorithm}[H]
	\caption{Quantum Spatial Best-Arm Identification}\label{qsbai:alg}
	\begin{algorithmic}[1]    
	\REQUIRE Graph $G=(V,\,A)$,\\
		Set of possible environment states $\Sigma$,\\
		Environment-state distribution $\eta_v:\Sigma\to [0,\,1]$ for each $v\in V$,\\
		Classical oracle $f:V\times \Sigma\to \{0,\,1\}$
		%Initial state $\ket{\Phi}\in\mathcal{H}_A$, Coin operator $C$, Quantum oracle $R_f$
	\ENSURE Selected arm $u_T$\vrule height0pt width0pt depth5pt
	\STATE $\ket{\Phi}\leftarrow \displaystyle\sum_{\substack{v\in V\\ \sigma\in\Sigma}} \sum_{\substack{v'\in \mathcal{N}_G(v)\\ \sigma'\in\Sigma}} \sqrt{\eta_v(\sigma)\eta_{v'}(\sigma')/|A|} \ket{v,\,\sigma;\,v',\,\sigma'}$\label{qsbai:alg:Phi}
	\FOR{$(v,\,\sigma)\in V\times\Sigma$}
	\FOR{$(v',\,\sigma')\in \mathcal{N}_G(v)\times\Sigma$}
	\FOR{$(v'',\,\sigma'')\in \mathcal{N}_G(v')\times\Sigma$}
	\STATE $\widetilde{a}\leftarrow ((v,\,\sigma),\,(v',\,\sigma')),\ \widetilde{b}\leftarrow ((v'',\,\sigma''),\,(v',\,\sigma'))$
	\STATE $\gamma_{\tilde{a}\tilde{b}} \leftarrow 2\sqrt{\eta_v(\sigma)\eta_{v''}(\sigma'')}/\deg_G(v')-\delta_{vv''}\delta_{\sigma\sigma''}$
	\ENDFOR
	\ENDFOR
	\ENDFOR
	\STATE $\displaystyle U_0 \leftarrow \sum_{\tilde{a}\in \tilde{A}}\sum_{\tilde{b}\in \mathcal{T}(\terminus(\tilde{a}))}\gamma'_{\tilde{a}\tilde{b}}\ketbra{\widetilde{b}^{-1}}{\widetilde{a}}$
	\STATE $R_f \leftarrow \displaystyle \sum_{\substack{v\in V\\ \sigma\in\Sigma}} \sum_{\substack{v'\in \mathcal{N}_G(v)\\ \sigma'\in\Sigma}}(-1)^{f(v',\sigma')}\ketbra{v,\,\sigma;\,v',\,\sigma'}{v,\,\sigma;\,v',\,\sigma'}$\label{qsbai:alg:Rf}
	\STATE $\ket{\Psi_0}\leftarrow \ket{\Phi}$\label{qsbai:alg:init}
%	\STATE $\widetilde{A} \leftarrow \{((v,\,\sigma),\,(v',\,\sigma'))\,|\,(v,\,v')\in A;\,\sigma,\,\sigma'\in \Sigma\}$
	\STATE $U\leftarrow U_0R_f$\label{qsbai:alg:tev}
	\FOR{$t\leftarrow 1$ \textbf{to} $T$}
	\STATE $\ket{\Psi_t} = U\ket{\Psi_{t-1}}$
	\ENDFOR\label{qsbai:alg:endtev}
	\FOR{$w\in V$}\label{qsbai:alg:beginP}
	\STATE $P_T(w) \leftarrow \displaystyle\sum_{\sigma'\in \Sigma}\sum_{v\in \mathcal{N}_G(w)}\sum_{\sigma\in \Sigma}|\braket{v,\,\sigma;\,w,\,\sigma'|{U}^t|\Phi}|^2$ 
	\ENDFOR \label{qsbai:alg:endP}
	\STATE Obtain $u_T \sim P_T$
	\RETURN $u_T$
	\end{algorithmic}
\end{algorithm}
\end{figure}

Here, we consider the state transition to describe the classical decision-making, which is crucial to construct the Szegedy's walk.
Let the selected arm at time step $t-1$ be $v\in V$, and the agent is making the $t$-th decision.
Here, assume that the environment state at time $t-1$ is $\sigma\in \Sigma$.
Then, the next arm is selected from the neighbors of $v$.
Assuming that it is uniformly selected, then the probability can be described by $1/\deg_G(v)$, where $\deg_G(v)$ is the degree of the vertex $v$.
After selecting the next arm, which is denoted by $v'$, the environment state is probabilistically determined following the probability distribution $\eta_{v'}$.
Therefore, the probability of selecting arm $v'$ and changing the environment state to $\sigma'$ turned out to be $\eta_{v'}(\sigma')/\deg_G(v)$.

\begin{figure}[t]
	\centering
	\includegraphics[width=68.4mm]{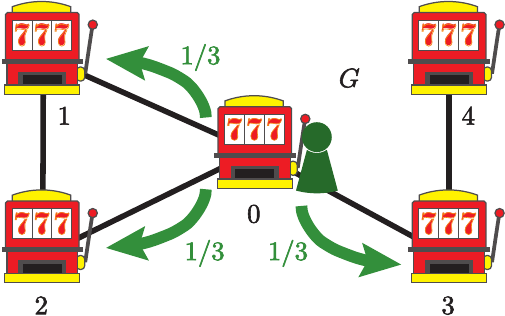}
	\caption{Example of arm selection on a graph $G = (V,\,A)$ with $V = \{0,\,1,\,2,\,3,\,4\}$ and $A = \{a,\,a^{-1}\,|\,a\in \{(0,\,1),\,(0,\,2),\,(0,\,3),\,(1,\,2),\,(3,\,4)\}\}$.
    Suppose the last selected arm is $0$.
    The next arm is then chosen uniformly at random from its three neighboring arms, $1$, $2$, and $3$, each with probability $1/3$.
    Since arm $4$ is not adjacent to arm $0$, the agent cannot select it in the next decision.}\label{qsbai:fig:example}
\end{figure}

\begin{figure}[t]
	\includegraphics[width=68.4mm]{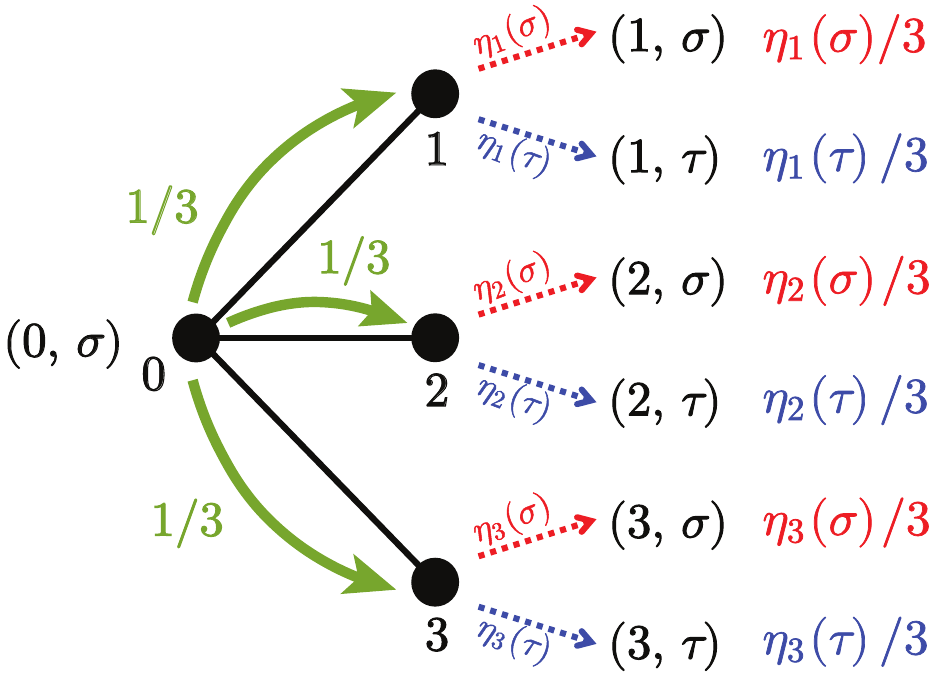}
	\caption{Possible state transitions in arm selection and environment states, corresponding to Fig.~\ref{qsbai:fig:example}.  
Each state is represented by a pair $(v,\,\sigma)$, where $v \in V$ is the selected arm and $\sigma \in \Sigma = \{\sigma,\,\tau\}$ is the environment state.  
Suppose the current state is $(0,\,\sigma)$.  
The next arm is chosen uniformly at random from its three neighbors $1$, $2$, and $3$, each with probability $1/3$, as shown in Fig.~\ref{qsbai:fig:example}.  
Afterward, the subsequent environment state is determined according to the distribution $\eta_v$ associated with the selected arm $v \in V$.  
For example, the probability of transitioning from $(0,\,\sigma)$ to $(1,\,\sigma)$ is the product of the arm-selection probability $1/3$ and the environment-transition probability $\eta_1(\sigma)$.  
The probabilities of other transitions are determined in the same manner.}\label{qsbai:fig:transition}
\end{figure}

Figure~\ref{qsbai:fig:example} illustrates an example of arm selection on a specific graph.
At this moment, the agent has just pulled the arm at vertex 0, and the next arms it can pull are 1, 2, or 3.
Each of these arms is selected with an equal probability of $1/3$.
Furthermore, the state transitions, including the environmental states and their transition probabilities, can be summarized as shown in Fig.~\ref{qsbai:fig:transition}.
This figure represents the case where there are two possible environmental states, $\sigma$ and $\tau$ (i.e., $\Sigma = \{\sigma,\,\tau\}$).

From this observation, we see that a state transition in this scheme is described by pairs consisting of the selected arm and the environment state. 
The above example is thus interpreted as the transition from $(v,\,\sigma)$ to $(v',\,\sigma')$.

\begin{figure*}[t]
	\centering
	\includegraphics[width=135mm]{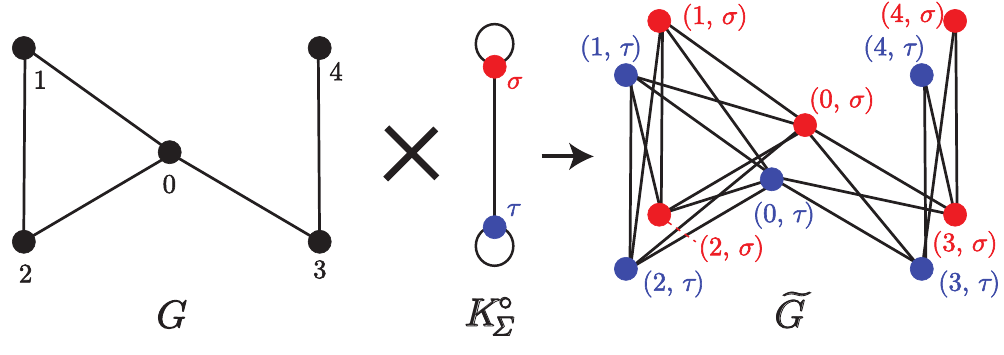}
	\caption{Example of constructing the executive graph $\widetilde{G}$ for the graph $G$ shown in Figs.~\ref{qsbai:fig:example} and \ref{qsbai:fig:transition}.  
The executive graph $\widetilde{G}$ represents possible state transitions in arm selection on $G$, where each state is expressed as a pair consisting of a selected arm and an environment state.  
Assuming that the environment state at each time step is determined independently, $\widetilde{G}$ is formalized as the direct product of the graph $G$ and the complete graph with self-loops indexed by the set $\Sigma$ of possible environment states, see Eq.~\eqref{qsbai:eq:Gtilde}.}\label{qsbai:fig:graph}
\end{figure*}

Next, we construct the \textit{executive graph} $\widetilde{G}$ for spatial BAI on graph $G$.
Based on the above discussion, each vertex of $\widetilde{G}$ should be labeled by a pair of an arm and an environment state; thus, the vertex set of $\widetilde{G}$ is defined as $V\times \Sigma$.
To describe the transition from vertex $v$ to a neighbor $v'$ in $G$, the pairs $(v,\,\sigma)$ and $(v',\,\sigma')$ must also be connected. 
Conversely, to enforce the spatial constraint given by the structure of $G$, pairs $(v,\,\sigma)$ and $(v'',\,\sigma'')$ must be disconnected if $v$ and $v''$ are not neighbors in $G$.
Since the variation of the environment state is independent of its previous value, the environment state does not impose additional disconnection conditions in $\widetilde{G}$.
In summary, the executive graph $\widetilde{G}$ is defined as
\begin{align}\label{qsbai:eq:Gtilde}
	\widetilde{G} := G\times K_{\Sigma}^\circ = \left(V\times \Sigma,\,\widetilde{A}\right),
\end{align}
where
\begin{align}
	\widetilde{A} := \{((v,\,\sigma),\,(v',\,\sigma'))\,|\,(v,\,v')\in A,\ \sigma,\,\sigma' \in \Sigma\}.
\end{align}
Here, $K_\Sigma^\circ := (\Sigma,\,\Sigma^2)$ is the complete graph with self-loops, whose vertices are labeled by environment states $\sigma\in \Sigma$. 
For two graphs $G_1$ and $G_2$, $G_1\times G_2$ denotes their direct product \cite{knuth2011art}.
For example, for the graph $G$ in Fig.~\ref{qsbai:fig:example} with two possible environment states $\Sigma = \{\sigma,\,\tau\}$, the executive graph $\widetilde{G}$ is constructed as shown in Fig.~\ref{qsbai:fig:graph}.

We formalize the state transitions for decision-making using a time-homogeneous random walk.
Let the transition probability from state $(v,\,\sigma)$ to $(v',\,\sigma')$ be
\begin{equation}\begin{split}
    p^{(v,\sigma)}_{(v',\sigma')} &:= p((v,\,\sigma),\,(v',\,\sigma')) \\
    &= 
    \begin{cases} 
    \dfrac{\eta_{v'}(\sigma')}{\deg_G(v)} & \text{if } v' \in \mathcal{N}_G(v), \\[8pt]
    0  & \text{if } v'\notin \mathcal{N}_G(v),
    \end{cases} 
\end{split}\label{qsbai:eq:p}\end{equation}
where $\mathcal{N}_G(v)$ denotes the set of neighbors of $v$ in $G$:
\begin{equation}
    \mathcal{N}_G(v) := \{v'\in V\,|\,(v,\,v')\in A\}.
\end{equation}
The well-definedness of $p^{(v,\sigma)}_{(v',\sigma')}$ is guaranteed by the normalization of $\eta_v$ for all $v\in V$.

Next, we define a function $\pi: V\times \Sigma\to \mathbb{R}$ as
\begin{equation}\label{qsbai:eq:pi}
    \pi(v,\,\sigma) = \frac{\deg_G(v)\eta_v(\sigma)}{|A|}.
\end{equation}
Here, $A$ is the set of arcs of the original graph $G$ (not of $\widetilde{G}$). 
Clearly, $\pi(v,\,\sigma)\geq 0$ for all $(v,\,\sigma)\in V\times \Sigma$, and $\pi$ is a valid probability distribution on $V\times \Sigma$, ensured by the normalization of $\eta_v$ and the relation $\sum_{v\in V}\deg_G(v)=|A|$. 
Moreover, the detailed balance condition holds:
\begin{equation}\label{qsbai:eq:DBC}
    m^{(v,\sigma)}_{(v',\sigma')} = \pi(v,\,\sigma)\cdot p^{(v,\sigma)}_{(v',\sigma')} 
    = \pi(v',\,\sigma')\cdot p^{(v',\sigma')}_{(v,\sigma)}.
\end{equation}
Thus, the function $\pi$ defined in Eq.~\eqref{qsbai:eq:pi} is the reversible distribution for the random walk associated with the transition probabilities in Eq.~\eqref{qsbai:eq:p} on the executive graph $\widetilde{G}$.

We now construct Szegedy’s walk using the transition probabilities and the reversible distribution defined above.
The system of QWs is given by the Hilbert space $\mathcal{H}_{\widetilde{A}}$:
\begin{equation}
    \mathcal{H}_{\widetilde{A}} = \spn\left\{ \ket{v,\,\sigma;\,v',\,\sigma'}\,\middle|\,((v,\,\sigma),\,(v',\,\sigma'))\in \widetilde{A}\right\}.
\end{equation}
We define the state $\ket{\Phi}$ on $\mathcal{H}_{\widetilde{A}}$ as
\begin{equation}
\begin{split}\label{qsbai:eq:Phi}
    \ket{\Phi} :=& \sum_{\substack{v\in V\\ \sigma\in\Sigma}} 
    \sum_{\substack{v'\in \mathcal{N}_G(v)\\ \sigma'\in\Sigma}} 
    \sqrt{m^{(v,\sigma)}_{(v',\sigma')}} \ket{v,\,\sigma;\,v',\,\sigma'} \\
       =& \sum_{\substack{v\in V\\ \sigma\in\Sigma}} 
       \sum_{\substack{v'\in \mathcal{N}_G(v)\\ \sigma'\in\Sigma}} 
       \sqrt{\frac{\eta_v(\sigma)\eta_{v'}(\sigma')}{|A|}} \ket{v,\,\sigma;\,v',\,\sigma'}.
\end{split}
\end{equation}
By using the normalization of $\eta_v$ and the relation $\sum_{v\in V}\deg_G(v)=|A|$, it follows that $\ket{\Phi}$ is a unit vector.

We next define the time evolution operator $U_0$ on $\mathcal{H}_{\widetilde{A}}$ as
\begin{equation}
    U_0 = \sum_{\tilde{a}\in \tilde{A}}\sum_{\tilde{b}\in \mathcal{T}(\terminus(\tilde{a}))}\gamma_{\tilde{a}\tilde{b}}'\ketbra{\tilde{b}^{-1}}{\tilde{a}}.
\end{equation}
For $\widetilde{a}=((v,\,\sigma),\,(v',\,\sigma'))$ and $\widetilde{b}=((v'',\,\sigma''),\,(v',\,\sigma'))$, the coefficient $\gamma_{\tilde{a}\tilde{b}}'$ is given by
\begin{equation}\label{qsbai:eq:gamma}
	\gamma_{\tilde{a}\tilde{b}}' = \frac{2\sqrt{\eta_v(\sigma)\eta_{v''}(\sigma'')}}{\deg_G(v')} - \delta_{vv''}\delta_{\sigma\sigma''},
\end{equation}
which follows from substituting Eq.~\eqref{qsbai:eq:p} into Eq.~\eqref{qsww:eq:gamma}.
Applying this formula to the definition of the time evolution operator of QWs (Eq.~\eqref{qwss:eq:U}), we obtain the time evolution operator for Szegedy’s walk.

\begin{figure}
    \centering
    \includegraphics[width=68.5mm]{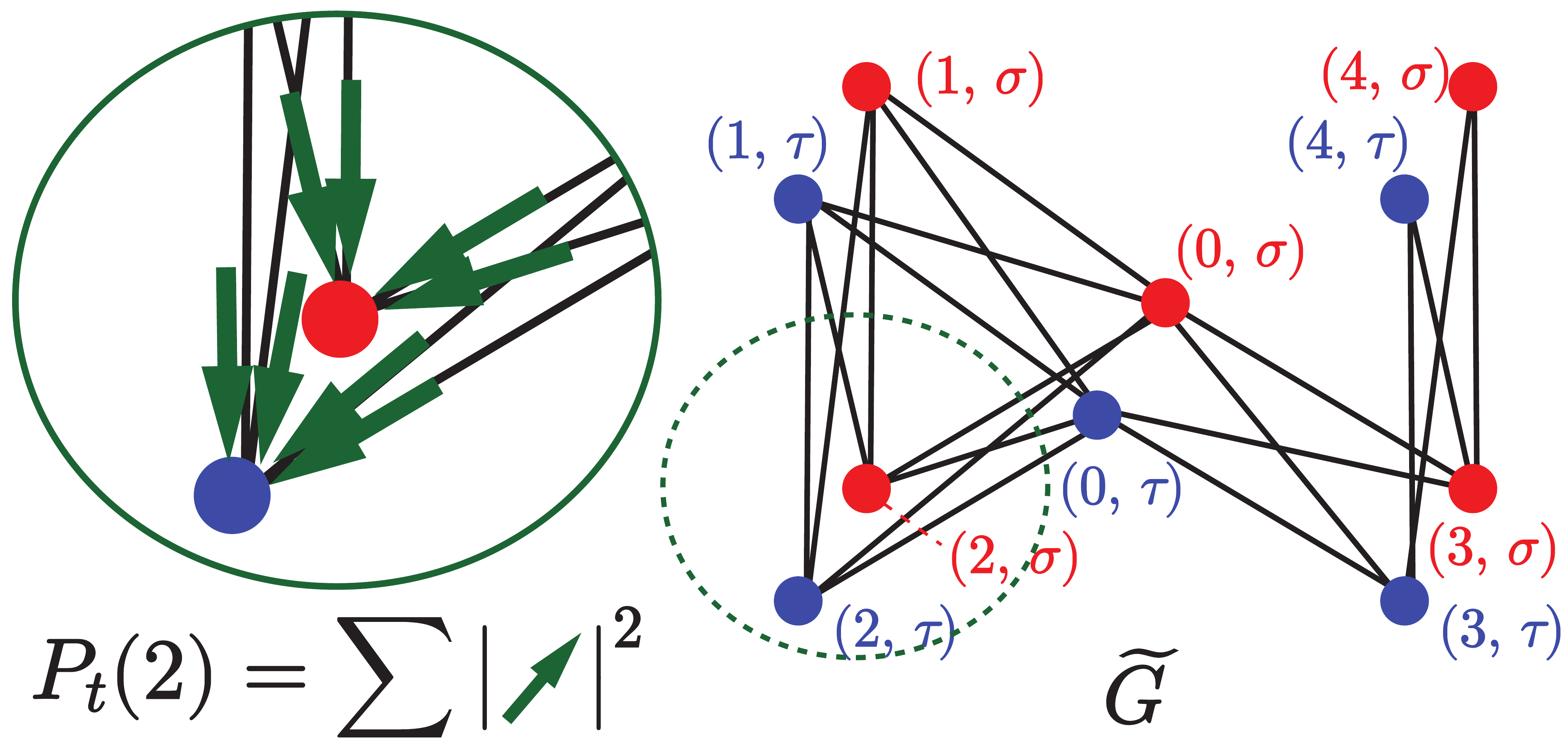}
    \caption{Example of the recommendation probability $P_t(w)$ given by Eq.~\eqref{qsbai:eq:recommendation} for $w=2$ on the graph $G$ in Fig.~\ref{qsbai:fig:example}.  
The recommendation probability $P_t(2)$ is calculated as the sum of the measurement probabilities of the states on the executive graph $\widetilde{G}$ (see Fig.~\ref{qsbai:fig:graph}) that involve arm $2$, regardless of the environment state.  By definition of the recommendation probability~\eqref{qwss:eq:measurement}, $P_t(2)$ is obtained by summing the squared absolute values of the probability amplitudes corresponding to arcs terminating at the relevant states.}\label{qsbai:fig:recommendation}
\end{figure}

We then construct the operator $R_f$ to mark the winning cases:
\begin{align}	
    R_f =& \sum_{\tilde{a}\in\tilde{A}}(-1)^{f(\terminus(\tilde{a}))}\ketbra{\tilde{a}}{\tilde{a}}\\
        =&\sum_{\substack{v\in V\\ \sigma\in\Sigma}} \sum_{\substack{v'\in \mathcal{N}_G(v)\\ \sigma'\in\Sigma}}(-1)^{f(v',\sigma')}\ketbra{v,\,\sigma;\,v',\,\sigma'}{v,\,\sigma;\,v',\,\sigma'}.\nonumber
\end{align}
The overall time evolution operator $U$ for QSBAI is then
\begin{equation}
    \begin{split}
        U &= U_0R_f\\
        &= \sum_{\tilde{a}\in\tilde{A}}\sum_{\tilde{b}\in\mathcal{T}(\terminus(\tilde{a}))}(-1)^{f(\terminus(\tilde{a}))}\gamma_{\tilde{a}\tilde{b}}'\ketbra{\tilde{b}^{-1}}{\tilde{a}}.
    \end{split}
\end{equation}
With this, we describe the sequence of states $(\ket{\Psi_t}\,|\,t\in \mathbb{N}_0)$ on $\mathcal{H}_{\widetilde{A}}$ as
\begin{equation}\label{qsbai:eq:Psi}
\begin{split}
    \ket{\Psi_0} &= \ket{\Phi},\\
    \ket{\Psi_t} &= {U}^t \ket{\Psi_{0}}\quad \text{for $t\geq 1$}.
\end{split}
\end{equation}
Finally, the recommendation probability $P_t(w)$ of arm $w$ after $t$ steps is defined by
\begin{align}
    P_t(w) &=\sum_{\sigma'\in \Sigma}\sum_{\substack{v\in \mathcal{N}_G(w)\\ \sigma\in \Sigma}}|\braket{v,\,\sigma;\,w,\,\sigma'|{U}^t|\Phi}|^2. \label{qsbai:eq:recommendation}
\end{align}
Here, $P_t(w)$ is defined by the sum of the probabilities that the arm associated with the measured state is $w$, regardless the environment state, as shown in Fig.~\ref{qsbai:fig:recommendation}.
The QSBAI algorithm aims to maximize this recommendation probability of the best arm $v^\star$.

%%%%%%%%%%%%%%%
%%%%%%%%%%%%%%%

\section{\label{exampleC}Example for complete graphs}%\setcounter{equation}{0}
This section presents an example of QSBAI on complete graphs with self loops.
Complete graphs have directed edges between all the pairs, including the cases both entries are the same.
It indicates that the case is essentially equivalent to the one without any spatial constraint.
That is, it is desired that the performance of QSBAI is what corresponds to the one of the original QBAI algorithm~\cite{casale2020quantum}.
%an examples: one where the algorithm is applied to a complete graph, and the other where it is applied to a complete bipartite graph.
%%%%
%%%%
%%%%
%\subsection{QSBAI on complete graphs}\label{sec5:CG}

%We first consider the case where the search space $G$ is a complete graph with self-loops. 
The complete graph $K_V^\circ = (V,\,V^2)$ for a given set $V$ contains connections between every pair of vertices, allowing a vertex to be connected to itself.  
Equivalently, the adjacency matrix $M_{K_V^\circ}$ of the complete graph is the $|V|\times |V|$ all-ones matrix $J_{|V|}$. 

Now, let us consider the adjacency matrix of the executive graph $\widetilde{K}_{V}^\circ = K_{V}^\circ\times K_\Sigma^\circ$. 
It is known that the adjacency matrix of a graph described by the direct product of two graphs can be represented as the tensor product of the adjacency matrices of the two.  
In this case, the adjacency matrix $M_{\widetilde{K}_{V}^\circ}$ becomes
\begin{equation}
	M_{\widetilde{K}_{V}^\circ} = M_{K_V^\circ} \otimes M_{K_\Sigma^\circ}.
\end{equation}
Since $M_{K_\Sigma^\circ}$ is the $|\Sigma|\times |\Sigma|$ all-ones matrix $J_{|\Sigma|}$, we obtain
\begin{equation}
	M_{\widetilde{K}_{V}^\circ} = J_{V\times \Sigma}.
\end{equation}
That is, $M_{\widetilde{K}_{V}^\circ}$ is also an all-ones matrix of size $|V\times \Sigma| \times |V\times \Sigma|$, which indicates that the executive graph $\widetilde{K}_{V}^\circ$ is itself a complete graph with self-loops indexed by $V\times \Sigma$. Thus, $\widetilde{K}_{V}^\circ$ imposes no spatial constraints.

In this setting, the following statement holds:
\begin{thm}\sl\label{exampleC:thm:complete}
Let
\begin{equation}\label{exampleC:eq:qbar_th}
	\overline{q} = \sum_{v\in V}\fraction{q_v}{|V|}\quad\text{and}\quad\theta = \arcsin(\sqrt{\overline{q}}).
\end{equation}
Then, for $s = \lfloor\pi /4\theta \rfloor$, which implies $s = \mathrm{O}(1/\sqrt{\overline{q}})$, the recommendation probability for the best arm $v^*$ at time $t$, denoted by $P_{t}(v^*)$, is maximized at $t=2s$, and satisfies
\begin{equation}\label{exampleC:eq:thm1}
        P_{2s}(v^*) \geq \frac{1}{|V|}\left(1 +\frac{(q_{v^*}-\overline{q})(1-2\overline{q})}{\overline{q}(1-\overline{q})}\right).
\end{equation}
\end{thm}

The detailed derivation of this result is given in Appendix~\ref{app:complete}, where arguments similar to those in Casal\'e et al.~\cite{casale2020quantum} on QBAI and Portugal~\cite{portugal2013quantum} on quantum search on complete graphs are presented.

\begin{figure}[t]
    \begin{center}
        \includegraphics[width=70mm]{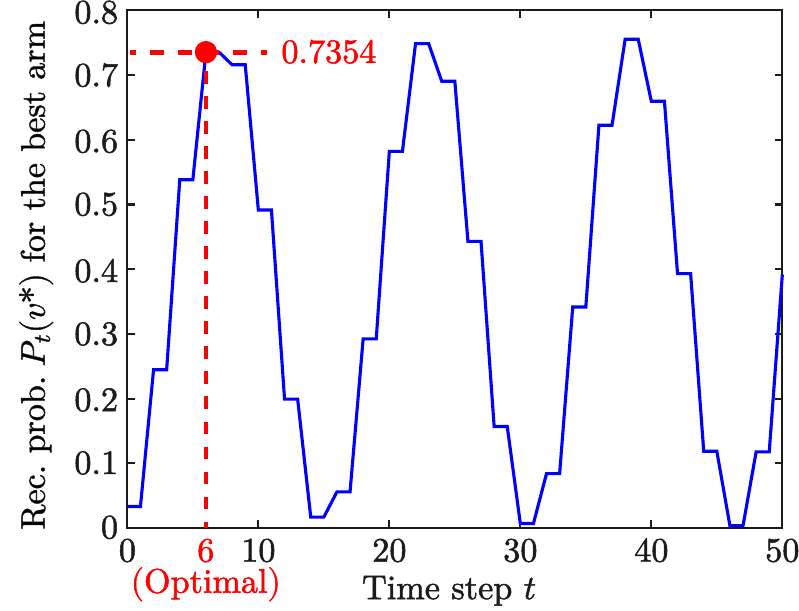}
        \caption{Variation of the recommendation probability $P_t(v^*)$ of the best arm $v^*$ over the time step $t$ for QSBAI on a complete graph with 30 vertices.}\label{exampleC:fig:complete}
    \end{center}
\end{figure}

Figure~\ref{exampleC:fig:complete} illustrates an example of QSBAI on a complete graph under the following conditions:
\begin{itemize}
	\setlength{\leftskip}{0.5cm}
	\item The set of vertices $V$ satisfies $|V| = 30$.
	\item The set of environments is $\Sigma = \{\sigma,\,\tau\}$.
	\item For any arm $v\in V$, the classical oracle $f$ is defined by
	\begin{equation}
		f(v,\,\sigma) = 1,\quad f(v,\,\tau) = 0,
	\end{equation}
	indicating that environment state $\sigma$ (resp. $\tau$) corresponds to the winning (resp. losing) case.
        \item For the best arm $v^*$, $\eta_{v^*}(\sigma)$ is set to $0.9$, while for any $v\in V\backslash \{v^*\}$, $\eta_v(\sigma)$ is set to $0.01$. 
        Since $\eta_v(\sigma)$ equals the winning probability $q_v$, this ensures that $v^*$ is the arm to be recommended eventually.
    \end{itemize}
The recommendation probability exhibits periodic behavior and attains its first maximum at $t=6$, with value $P_6(v^*) \simeq 0.7354$. 
The right-hand side of inequality~\eqref{exampleC:eq:thm1} evaluates to $0.7264$ in this setting, verifying the bound.

This result coincides with that obtained by $s$ steps of the previous QBAI approach presented in Ref.~\cite{casale2020quantum}. In a complete graph with self-loops, there are essentially no spatial constraints, which aligns this case with the non-spatial BAI setting.

%%%%%%%%%%%%%%%
%%%%%%%%%%%%%%%

\section{\label{exampleBP}Example for complete bipartite graphs}%\setcounter{equation}{0}
We consider an example of QSBAI on complete bipartite graphs.
Complete bipartite graphs have two clusters; each cluster has one or more vertices all of which are connected with the ones on the other cluster but not with the ones on the same cluster.
\red{This graph class captures a basic form of structured accessibility: the action space is divided into two clusters, and admissible transitions are determined by that partition. Such a structure provides spatially-constrained decision-making problems, in which the agent cannot move arbitrarily among all candidate actions. Complete bipartite graphs thus serve as a suitable model for examining how graph-constrained accessibility influences the best-arm identification, and allow rigorous analysis of how this structure modifies the behavior of QSBAI, compared to the case without spatial constraints.}

Let two clusters of a complete bipartite graph be indexed by two sets $V_1$ and $V_2$; specifically, the graph $K_{V_1,V_2} = (V_1\cup V_2,\,A)$, where the arc set $A$ is given by
\begin{equation}
	A = (V_1\times V_2)\cup (V_2\times V_1).
\end{equation}
This means that vertices are divided into two clusters, $V_1$ and $V_2$. Probability amplitudes can flow between vertices in different clusters, but not between vertices within the same cluster.

The adjacency matrix of $K_{V_1,V_2}$ satisfies
\begin{equation}
    M_{K_{V_1,V_2}} \simeq J_{|V_1|,|V_2|}\oplus J_{|V_2|,|V_1|} \simeq \twobytwo{J_{|V_1|,|V_2|}}{O_{|V_1|}}{O_{|V_2|}}{J_{|V_2|,|V_1|}},
\end{equation}
where $J_{r,c}$ is the $r\times c$ all-ones matrix and $O_{m}$ is the $m\times m$ zero matrix. 
Thus, the adjacency matrix of the executive graph $\widetilde{K}_{V_1,V_2}$ is
\begin{equation}\begin{split}
    M_{\widetilde{K}_{V_1,V_2}} &\simeq (J_{|V_1|,|V_2|}\oplus J_{|V_2|,|V_1|})\otimes J_{|\Sigma|}\\
        &= J_{|V_1\times \Sigma|,|V_2\times \Sigma|}\oplus J_{|V_2\times \Sigma|,|V_1\times \Sigma|}.
\end{split}\end{equation}
Therefore, the executive graph $\widetilde{K}_{V_1,V_2}$ of QSBAI on a complete bipartite graph $K_{V_1,V_2}$ is also a complete bipartite graph, with its two clusters indexed by $V_1\times \Sigma$ and $V_2\times \Sigma$, respectively.

In this setting, the following statement holds:
\begin{thm}\sl\label{exampleBP:thm:complete_bi}
	Assume that the best arm $v^*$ belongs to $V_i\,(i =1,\,2)$. Define
	\begin{equation}\label{exampleBP:eq:qbari_th}
		\overline{q}_i = \sum_{v\in V_i}\frac{q_v}{|V_i|}\quad\text{and}\quad \theta_i = \arcsin(\sqrt{\overline{q}_i}).
	\end{equation}
	Then, for $s = \lfloor\pi/4\theta_i\rfloor$, which implies $s = \mathrm{O}(1/\sqrt{\overline{q}_i})$, the recommendation probability for the best arm $v^*$ at time $t$, denoted by $P_t(v^*)$, is maximized at $t=2s$, and satisfies
	\begin{equation}\label{exampleBP:eq:thm2}
            P_{2s}(v^*) \geq \frac{1}{2|V_i|}\left(1 +\frac{(q_{v^*}-\overline{q}_i)(1-2\overline{q}_i)}{\overline{q}_i(1-\overline{q}_i)}\right).
	\end{equation}
\end{thm}

The detailed derivation of this result is deferred to Appendix~\ref{app:complete_bi}, where arguments similar to those in Rhodes and Wong~\cite{rhodes2019quantum} on quantum search on complete bipartite graphs are presented.

\begin{figure}[t]
    \begin{center}
        \includegraphics[width=70mm]{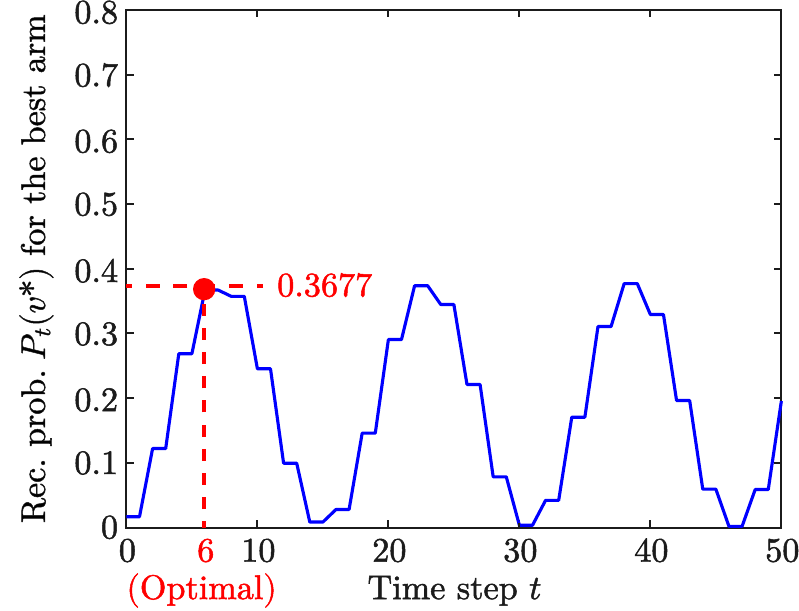}
        \caption{Variation of the recommendation probability $P_t(v^*)$ of the best arm $v^*$ over the time step $t$ for QSBAI on a complete bipartite graph with 30 vertices on one side and 10 vertices on the other.}\label{exampleBP:fig:K30_10}
    \end{center}
\end{figure}

Figure~\ref{exampleBP:fig:K30_10} illustrates an example of QSBAI on a complete bipartite graph under the following conditions:
\begin{itemize}
	\setlength{\leftskip}{0.5cm}
	\item The sets of vertices satisfy $|V_1| = 30$ and $|V_2| = 10$.
	\item The set of environments is $\Sigma = \{\sigma,\,\tau\}$.
	\item For any arm $v\in V$, the classical oracle $f$ is defined by
	\begin{equation}
		f(v,\,\sigma) = 1,\quad f(v,\,\tau) = 0,
	\end{equation}
	indicating that environment state $\sigma$ (resp. $\tau$) corresponds to the winning (resp. losing) case.
        \item For the best arm $v^*$, $\eta_{v^*}(\sigma)$ is set to $0.9$, while for any $v\in V\backslash \{v^*\}$, $\eta_v(\sigma)$ is set to $0.01$.
        Since $\eta_v(\sigma)$ equals the winning probability $q_v$, this ensures that $v^*$ is the arm to be recommended eventually.
\end{itemize}
The recommendation probability again exhibits periodic behavior and reaches its first maximum at $t=6$, with value $P_6(v^*) \simeq 0.3677$. 
The right-hand side of inequality~\eqref{exampleBP:eq:thm2} evaluates to $0.3632$ in this setting, verifying the bound.

It should be noted that the optimized recommendation probability in this example is half as large as that in the non-spatial QBAI case for arms in the cluster containing the best arm. However, the order of time steps required to reach the optimized recommendation probability coincides with that in the non-spatial QBAI setting. 
\red{This contrast indicates that the bipartite accessibility structure has a visible effect on QSBAI. 
Compared with the non-spatial case, the achievable recommendation probability is reduced 
while the order of the optimal time step is preserved. 
The appearance of the cluster-dependent quantities $q_i$ and $|V_i|$ also shows that 
the performance of QSBAI is affected by the bipartite structure itself.}

%%%%%%%%%%%%%%%
%%%%%%%%%%%%%%%

\section{\label{discussion}Discussions}%\setcounter{equation}{0}
Further tasks are required to render the results in this paper practically applicable. 
First, we cannot yet claim that our proposed method outperforms classical models that address BAI with spatial constraints. 
One reason is that the optimal time step $t_0$ depends on information inaccessible to the agent, as highlighted in a previous study \cite{casale2020quantum}. 
Indeed, as shown in Eq.~\eqref{exampleC:eq:qbar_th}, the optimal time step $t_0 = 2s$ is expressed by $\theta = \arcsin \sqrt{\overline{q}}$, which depends on the winning probabilities of the arms defined by the probability distribution $\eta_v$, unknown to the agent.
Currently, knowledge of $\eta_v$ is required to optimize the number of time-evolution steps; however, if $\eta_v$ were known, the best arm would already be apparent. 
\red{This issue is related to the broader problem of determining an appropriate amplification length when the effective amplification parameter is unknown. The environment-state distributions in QSBAI affect the construction of the initial state and the walk dynamics themselves. The estimation-based strategies~\cite{brassard2002quantum} and the robust amplification schemes~\cite{grover2005fixed,yoder2014fixed} are promising approaches for our future work, although their adaptation to QSBAI requires an additional reformulation.}

Another limitation is that it remains unclear how to construct the classical model that achieves the best performance for BAI with spatial constraints. 
In this paper, we focused on the case of a complete bipartite graph, where spatial structure imposes constraints. 
The consideration of arms arranged on a complete bipartite graph is itself novel. 
Therefore, further work is needed to explore which classical model can provide a fair benchmark for evaluating QSBAI. 
One simple idea for a spatial \textit{classical} BAI is to utilize the Markov chain on which the Szegedy walk in QSBAI is based, while another is to extend established BAI solutions, such as UCB-E \cite{audibert2010best}, to settings with spatial structures. 
To advance this study, we must investigate the problem from both quantum and classical perspectives. 

Next, we examined QSBAI on specific graphs. 
To claim that our method is broadly effective, it will be necessary to analyze this model on arbitrary graphs—specifically, to determine how time complexity is affected by spatial constraints and whether QSBAI consistently outperforms the corresponding classical models. 
In addition, extending the analysis to random graphs represents an open and interesting direction for future research. 
In that context, we must consider whether general results from conventional QW-based search can be applied to the analysis. 
This may be possible if QW-based search can be discussed on the directed product graph. 
\red{Another possible direction for future work is to explore QSBAI with weighted self loops, such as the recent studies of lackadaisical QWs on complete bipartite graph~\cite{rhodes2019search}, locally arc-transitive graph~\cite{hoyer2020analysis}, and vertex-transitive graph~\cite{rhodes2020search}.}

Finally, from the perspective of establishing a foundation for QW-based MABs, incorporating operations corresponding to exploitation into this algorithm is a crucial future task. 
If successful, such an extension would mark significant progress in the development of quantum bandit algorithms and quantum decision-making on networks.

%%%%%%%%%%%%%%%
%%%%%%%%%%%%%%%

\section{\label{conclusion}Conclusions}%\setcounter{equation}{0}
This study has introduced a novel quantum algorithm, called Quantum Spatial Best-Arm Identification (QSBAI), which leverages quantum walks to adapt traditional best-arm identification (BAI) to environments with spatial or network constraints. 
The key step was to formulate the arms as vertices in a graph so that restricted accessibility becomes part of the decision problem itself.
On top of this formulation, we constructed a quantum-walk-based search procedure to explore and identify the optimal arm.

The results demonstrated that, on specific types of graphs such as complete bipartite graphs, the quantum approach can achieve favorable probability amplification of the best arm within a certain number of steps, even considering the spatial limitations. 
\red{The analysis on complete bipartite graphs shows that a structured accessibility constraint can reduce the achievable recommendation probability, while preserving the order of the optimal time step.}
This indicates potential advantages in speeding up decision-making processes compared to classical algorithms, particularly in environments where actions are spatially constrained.

\red{The present work is a first theoretical step, rather than a complete practical solution.
While our analyses on complete and complete bipartite graphs have clarified fundamental properties of QSBAI, 
important issues still remain, such as the dependence of the optimal time-evolution length on unknown environment-state distributions and the design of fair classical benchmarks under spatial constraints.}

\red{This research lays foundational groundwork for quantum best-arm identification in graph-structured environments and for quantum-walk-based decision making under restricted accessibility. Our future works focus on extending the analyses to broader graph classes and making QSBAI more practical, particularly in choosing the time-evolution length without prior knowledge of the environment-state distributions.}

%However, the work is still in developmental stages. There remain open challenges, such as broader analysis on diverse graph structures, including arbitrary and random graphs, and establishing classical benchmarks for a fair comparison. Additionally, incorporating strategies for balancing exploration and exploitation—core to bandit algorithms—within a quantum framework is crucial for practical implementations.

%Overall, this research lays foundational groundwork for quantum algorithms tailored to complex, network-structured decision environments, opening new pathways for quantum-enhanced learning and adaptive decision-making on networks. 
%Future efforts should focus on expanding these models, 
%analyzing their complexities across different graph types, 
%\red{exploring weighted-self-loop extensions relevant to lackadaisical quantum walks,}
%and integrating classical ideas like exploration--exploitation trade-offs to realize a comprehensive quantum bandit framework.

%%%%%%%%%%%%%%%
%%%%%%%%%%%%%%%

\begin{acknowledgments}
This work was supported by the SPRING program (JPMJSP2108), CREST project (JPMJCR24R2) funded by the Japan Science and Technology Agency, and Grant-in-Aid for JSPS Fellows (JP23KJ0384), Transformative Research Areas (A) (JP22H05195, JP22H05197), Scientific Research (A) (JP25H01129), and Early-Career Scientists (JP25K21294) funded by the Japan Society for the Promotion of Science.\par
A major part of this study was conducted under supervision by Former Professor Makoto Naruse at the University of Tokyo, who passed away in September 2023. We would like to express our sincere gratitude and deepest condolences here.
\end{acknowledgments}

\appendix

\section{\label{app:complete}Proof of Theorem~\ref{exampleC:thm:complete}}
Here, we define the Hilbert space spanned by quantized vertices:
\begin{equation}
    \mathcal{H}_\rP := \spn \{\ket{v,\,\sigma}\,|\,(v,\,\sigma)\in V\times \Sigma\}.
\end{equation}
By setting $\ket{v,\,\sigma,\,v',\,\sigma'} = \ket{v,\,\sigma}\otimes \ket{v',\,\sigma'}$, we can treat $\mathcal{H}_{\widetilde{A}}$ as the compound Hilbert space comprising two $\mathcal{H}_\rP$-s; that is, $\mathcal{H}_{\widetilde{A}} = \mathcal{H}_\rP\otimes \mathcal{H}_\rP$. Since all the vertices are connected with each other including themselves, indicating $N_G(v) = V$ holds for all $v\in V$, Eqs.~\eqref{qsbai:eq:Phi} and \eqref{qsbai:eq:gamma} are rewritten as
\begin{equation}
    \ket{\Phi} = \ket{\varphi}\otimes \ket{\varphi}
\end{equation}
with
\begin{equation}
    \ket{\varphi} := \sum_{\substack{v\in V\\ \sigma\in\Sigma}} \sqrt{\fraction{\eta_v(\sigma)}{|V|}}\ket{v,\,\sigma}
\end{equation}
and
\begin{equation}
    \gamma_{\tilde{a}\tilde{b}}' = \fraction{2\sqrt{\eta_v(\sigma)\eta_{v''}(\sigma'')}}{|V|}-\delta_{vv''}\delta_{\sigma\sigma''},
\end{equation}
for $\widetilde{a} = ((v,\,\sigma),\,(v',\,\sigma'))$ and $\widetilde{b} = ((v'',\,\sigma''),\,(v',\,\sigma'))$.
Here, the unitary operator $U_0$ defined by
\begin{equation}
    U_0 = \sum_{\tilde{a}\in \tilde{A}}\sum_{\tilde{b}\in \mathcal{T}(\terminus(\tilde{a}))}\gamma_{\tilde{a}\tilde{b}}'\ketbra{\tilde{b}^{-1}}{\tilde{a}}
\end{equation}
constructing $U$ as $U =U_0R_f$, can be transformed as 
\begin{equation}
    U_0 = S(D_{\varphi}\otimes I_{\mathcal{H}_\rP})
\end{equation}
with
\begin{equation}\label{app:complete:eq:S}
    S = \sum_{\tilde{a}\in A}\ketbra{\tilde{a}^{-1}}{\tilde{a}}
\end{equation}
and
\begin{equation}
    D_\varphi := 2\ketbra{\varphi}{\varphi} -I_{\mathcal{H}_\rP}.
\end{equation}
It should be noted that $S$ corresponds to the flip-flop shift operator in terms of \textit{coined} quantum walks \cite{portugal2013quantum}.
Besides, the quantum oracle $R_f$ can be transformed into
\begin{equation}
    R_f = I_{\mathcal{H}_\rP}\otimes {R_{f}}_{\rP}
\end{equation}
with
\begin{equation}
    {R_{f}}_{\rP} = \sum_{\substack{v\in V\\ \sigma\in\Sigma}} (-1)^{f(v,\sigma)}\ketbra{v,\,\sigma}{v,\,\sigma}.
\end{equation}
Using these notations, $U$ is transformed as follows:
\begin{equation}
    U = S(D_{\varphi}\otimes {R_f}_\rP).
\end{equation}
By applying this $U$ to time evolution of $\ket{\Phi}$, we obtain the following: for a positive integer $s$,
\begin{align}
    U^{2s-1}\ket{\Phi} &= {R_f}_\rP(D_\varphi {R_f}_\rP)^{s-1}\ket{\varphi}\otimes (D_\varphi {R_f}_\rP)^{s-1}\ket{\varphi},\\
    U^{2s}\ket{\Phi} &= {R_f}_\rP(D_\varphi {R_f}_\rP)^{s-1}\ket{\varphi}\otimes (D_\varphi {R_f}_\rP)^{s}\ket{\varphi}.
\end{align}
Thus, the recommendation probability $P_{2s}(w)$ of arm $w$ after $2s$ steps of time evolution is calculated as
\begin{equation}\label{app:complete:eq:P}
\begin{split}
    P_{2s}(w) &= \sum_{\sigma'\in\Sigma}\sum_{\substack{v\in V\\ \sigma\in\Sigma}}|(\bra{v,\,\sigma}\otimes\bra{w,\,\sigma'})U^{2s}\ket{\Phi}|^2\\
    &=\sum_{\substack{v\in V\\ \sigma\in\Sigma}}|\braket{v,\,\sigma|{R_f}_\rP(D_\varphi {R_f}_\rP)^{s-1}|\varphi}|^2\hspace{-25ex}\raisebox{-20pt}{$\displaystyle\cdot\sum_{\sigma'\in\Sigma}|\braket{w,\,\sigma|(D_\varphi {R_f}_\rP)^{s}|\varphi}|^2$}\\
    &= \sum_{\sigma'\in\Sigma}|\braket{w,\,\sigma|(D_\varphi {R_f}_\rP)^{s}|\varphi}|^2.
\end{split}
\end{equation}
Let $\ket{\omega}$ and $\ket{\omega^\perp}$ be unit vectors given by
\begin{align}
    \ket{\omega} &= \fraction{1}{\sqrt{\overline{q}}}\sum_{(v,\sigma)\in W}\sqrt{\fraction{\eta_v(\sigma)}{|V|}}\ket{v,\,\sigma},\\
    \ket{\omega^\perp} &= \fraction{1}{\sqrt{1-\overline{q}}}\sum_{(v,\sigma)\in V\setminus W}\sqrt{\fraction{\eta_v(\sigma)}{|V|}}\ket{v,\,\sigma}.
\end{align}
Recall that $W$ is the subset of $V\times\Sigma$ that determines which pair of an arm and an environment state leads to generating a reward, see Eq.~\eqref{qsbai:eq:f}, and $\overline{q}$ is defined in Eq.~\eqref{exampleC:eq:qbar_th}, with which $\sqrt{\overline{q}}$ is represented by $\sin\theta$.
It should be noted that $\ket{\omega}$ and $\ket{\omega^\perp}$ are orthogonal with each other. 
Here $\ket{\varphi}$ can be rewritten by the linear combinations of $\ket{\omega}$ and $\ket{\omega^\perp}$ as follows:
\begin{equation}\begin{split}
    \ket{\varphi} &= \sqrt{\overline{q}}\ket{\omega} + \sqrt{1-\overline{q}}\ket{\omega^\perp}\\
    &= [\,\ket{\omega}\ \ \ket{\omega^\perp}\,] \ket{\widetilde{\varphi}}
\end{split}\end{equation}
with
\begin{equation}
    \ket{\widetilde{\varphi}} = \twovec{\sin\theta}{\cos\theta}.
\end{equation}
Besides, the representative matrix of operators $D_\varphi$ and ${R_f}_\rP$ for the basis $\{\ket{\omega},\,\ket{\omega^\perp}\}$ are obtained by $\widetilde{D}_\varphi$ and $\widetilde{R_f}_\rP$, given as follows, respectively:
\begin{align}
    \widetilde{D}_\varphi &= 2\ketbra{\widetilde{\varphi}}{\widetilde{\varphi}}-I_2 = \twobytwo[r]{-\cos 2\theta}{\sin 2\theta}{\sin 2\theta}{\cos 2\theta},\\
    \widetilde{R_f}_\rP &= \twobytwo[r]{-1}{0}{0}{1},
\end{align}
where $I_2$ is the $2\times 2$ identity matrix.
Therefore, $(D_\varphi {R_f}_\rP)^s\ket{\varphi}$ can be transformed into
\begin{equation}\begin{split}
    (D_\varphi {R_f}_\rP)^s\ket{\varphi} = [\ket{\omega}\ \,\ket{\omega^\perp}]\widetilde{D}_\varphi^s \widetilde{R_f}_\rP^s\twovec{\sin\theta}{\cos\theta}&\\
    = \sin((2s+1)\theta) \ket{\omega} +\cos ((2s+1)\theta) \ket{\omega^\perp}&.
\end{split}\end{equation}
Substituting this to Eq.~\eqref{app:complete:eq:P}, we have
\begin{equation}\label{app:complete:eq:P2}\begin{split}
	&\hspace{-1ex}P_{2s}(w)\\
	&=  \sum_{\sigma\in\Sigma^*_w}\sin^2((2s+1)\theta)|\braket{w,\,\sigma|\omega}|^2 \hspace{-30ex}\raisebox{-25pt}{$\displaystyle+ \sum_{\sigma\in\Sigma\setminus\Sigma^*_w}\cos^2((2s+1)\theta)|\braket{w,\,\sigma|\omega^\perp}|^2$}\\
	&=  \fraction{\sin^2((2s+1)\theta)}{\overline{q}|V|}\sum_{\sigma\in\Sigma^*_w}\eta_w(\sigma) \hspace{-25ex}\raisebox{-35pt}{$\mathrel{+}\fraction{\cos^2((2s+1)\theta)}{(1-\overline{q})|V|}\sum_{\sigma\in\Sigma\setminus\Sigma^*_w}\eta_w(\sigma)$}\\
	&= \fraction{q_w}{\overline{q}|V|}\sin^2((2s+1)\theta) +\fraction{1-q_w}{(1-\overline{q})|V|}\cos^2((2s+1)\theta)\\
	&= \fraction{1}{|V|}\left(1 +\fraction{(q_w-\overline{q})(\sin^2((2s+1)\theta)-\overline{q})}{\overline{q}(1-\overline{q})}\right).
\end{split}\end{equation}
It can be seen that $P_{2s}(w)$ is maximized when $s$ is set to be such that $\sin^2((2s+1)\theta)$ is closest to $1$, which indicates that
\begin{equation}
	s = s_0 := \left\lfloor\fraction{\pi}{4\theta}\right\rfloor.
\end{equation}
Herein, there exists $\delta\in [0,\,1)$ which satisfies
\begin{equation}
	s_0 = \fraction{\pi}{4\theta} -\delta.
\end{equation}
Using this $\delta$, we can transform $\sin((2s_0+1)\theta)$ as follows:
\begin{equation}\begin{split}
	\sin^2((2s_0+1)\theta) &= \sin^2\left( \left\{ 2\left(\fraction{\pi}{4\theta}-\delta\right) +1\right\}\theta\right)\\
	&= \sin^2\left(\left( \fraction{\pi}{2} +1-2\delta\right)\theta\right)\\
	&= \cos^2 ((1-2\delta)\theta)\\
	&= 1-\sin^2((1-2\delta)\theta).
\end{split}\end{equation}
By the relation $-\theta< (1-2\delta)\theta \leq\theta$, the inequality $|\sin((1-2\delta)\theta)|\leq |\sin\theta| = \sqrt{\overline{q}}$ holds. 
Thus, $\sin^2((2s_0+1)\theta)$ can be estimated as
\begin{equation}
	\sin^2((2s_0+1)\theta) \geq 1-\overline{q}.
\end{equation}
Applying this to Eq.~\eqref{app:complete:eq:P2} clarifies that the recommendation probability $P_{2s_0}(w)$ of arm $w$ at the optimal time $2s_0$ is estimated as
\begin{equation}
	P_{2s_0}(w) \geq \fraction{1}{|V|}\left(1 +\fraction{(q_w-\overline{q})(1-2\overline{q})}{\overline{q}(1-\overline{q})}\right),
\end{equation}
which for the best arm $v^*$ is the desired result.

\section{\label{app:complete_bi}Proof of Theorem~\ref{exampleBP:thm:complete_bi}}
Recalling that the vertex set $V$ is decomposed into two clusters $V_1$ and $V_2$, and all the vertices inhabiting a cluster are connected with all the ones inhabiting the other, $\ket{\Phi}$ can be rewritten as follows:
\begin{equation}
    \ket{\Phi} = \sum_{\substack{v\in V_1\\ \sigma\in\Sigma}}\sum_{\substack{v'\in V_2\\ \sigma'\in\Sigma}}\sqrt{\fraction{\eta_v(\sigma)\eta_{v'}(\sigma')}{2|V_1||V_2|}}\hspace{-15ex}\raisebox{-25pt}{$\left(\ket{v,\,\sigma;\,v',\,\sigma'} +\ket{v',\,\sigma';\,v,\,\sigma} \right),$}
\end{equation}
wherein the number of arcs in a complete bipartite digraph equals twice the product of the vertex counts in the two clusters.

Here, we define the following unit vectors:
\begin{align}
    \ket{\omega_{11}} &= \fraction{1}{c_{11}}\sum_{\substack{v\in V_1\\ \sigma\in\Sigma_v^*}}\sum_{\substack{v'\in V_2\\ \sigma'\in\Sigma_{v'}^*}}\sqrt{\fraction{\eta_{v}(\sigma)\eta_{v'}(\sigma')}{2|V_1||V_2|}}\ket{v,\,\sigma;\,v',\,\sigma'},\\
    \ket{\omega_{10}} &= \fraction{1}{c_{10}}\sum_{\substack{v\in V_1\\ \sigma\in\Sigma_v^*}}\sum_{\substack{v'\in V_2\\ \sigma'\not\in\Sigma_{v'}^*}}\sqrt{\fraction{\eta_{v}(\sigma)\eta_{v'}(\sigma')}{2|V_1||V_2|}}\ket{v,\,\sigma;\,v',\,\sigma'},
\end{align}
\begin{align}
    \ket{\omega_{01}} &= \fraction{1}{c_{01}}\sum_{\substack{v\in V_1\\ \sigma\not\in\Sigma_v^*}}\sum_{\substack{v'\in V_2\\ \sigma'\in\Sigma_{v'}^*}}\sqrt{\fraction{\eta_{v}(\sigma)\eta_{v'}(\sigma')}{2|V_1||V_2|}}\ket{v,\,\sigma;\,v',\,\sigma'},\\
    \ket{\omega_{00}} &= \fraction{1}{c_{00}}\sum_{\substack{v\in V_1\\ \sigma\not\in\Sigma_v^*}}\sum_{\substack{v'\in V_2\\ \sigma'\not\in\Sigma_{v'}^*}}\sqrt{\fraction{\eta_{v}(\sigma)\eta_{v'}(\sigma')}{2|V_1||V_2|}}\ket{v,\,\sigma;\,v',\,\sigma'},
\end{align}
where $c_{ij}$-s $(i,\,j\in\{0,\,1\})$ are defined as follows:
\begin{align}
    c_{11} = \sqrt{\sum_{\substack{v\in V_1\\ \sigma\in\Sigma_v^*}}\sum_{\substack{v'\in V_2\\ \sigma'\in\Sigma_{v'}^*}}\fraction{\eta_{v}(\sigma)\eta_{v'}(\sigma')}{2|V_1||V_2|}},\\
    c_{10} = \sqrt{\sum_{\substack{v\in V_1\\ \sigma\in\Sigma_v^*}}\sum_{\substack{v'\in V_2\\ \sigma'\not\in\Sigma_{v'}^*}}\fraction{\eta_{v}(\sigma)\eta_{v'}(\sigma')}{2|V_1||V_2|}},\\
    c_{01} = \sqrt{\sum_{\substack{v\in V_1\\ \sigma\not\in\Sigma_v^*}}\sum_{\substack{v'\in V_2\\ \sigma'\in\Sigma_{v'}^*}}\fraction{\eta_{v}(\sigma)\eta_{v'}(\sigma')}{2|V_1||V_2|}},\\
    c_{00} = \sqrt{\sum_{\substack{v\in V_1\\ \sigma\not\in\Sigma_v^*}}\sum_{\substack{v'\in V_2\\ \sigma'\not\in\Sigma_{v'}^*}}\fraction{\eta_{v}(\sigma)\eta_{v'}(\sigma')}{2|V_1||V_2|}}.
\end{align}
Using the notations $\overline{q}_i$ and $\theta_i$ introduced in \eqref{exampleBP:eq:qbari_th}, they are transformed into
\begin{align}
    c_{11} &= \sqrt{\fraction{1}{2}\overline{q}_1\overline{q}_2} = \fraction{1}{\sqrt{2}}\sin\theta_1 \sin\theta_2,\\
    c_{10} &= \sqrt{\fraction{1}{2}\overline{q}_1(1-\overline{q}_2)} = \fraction{1}{\sqrt{2}}\sin\theta_1 \cos\theta_2,\\
    c_{01} &= \sqrt{\fraction{1}{2}(1-\overline{q}_1)\overline{q}_2} = \fraction{1}{\sqrt{2}}\cos\theta_1 \sin\theta_2,\\
    c_{00} &= \sqrt{\fraction{1}{2}(1-\overline{q}_1)(1-\overline{q}_2)} = \fraction{1}{\sqrt{2}}\cos\theta_1 \cos\theta_2.
%    c_{11} &= \sqrt{\fraction{1}{2}(1-\overline{q}_1)(1-\overline{q}_2)} = \fraction{1}{\sqrt{2}}\cos\theta_1 \cos\theta_2.
\end{align}
These notations allow to rewrite $\ket{\Phi}$ as
\begin{align}\begin{split}
    \ket{\Phi} =& \fraction{1}{\sqrt{2}}\sin\theta_1\sin\theta_2 (\ket{\omega_{11}}+\ket{\omega_{11}^{-1}})\\
    & +\fraction{1}{\sqrt{2}}\sin\theta_1\cos\theta_2 (\ket{\omega_{10}}+\ket{\omega_{10}^{-1}})\\
    & +\fraction{1}{\sqrt{2}}\cos\theta_1\sin\theta_2 (\ket{\omega_{01}}+\ket{\omega_{01}^{-1}})\\
    & +\fraction{1}{\sqrt{2}}\cos\theta_1\cos\theta_2 (\ket{\omega_{00}}+\ket{\omega_{00}^{-1}})
\end{split}\\
    \phantom{\ket{\Phi}}=& [\ket{\omega_{11}}\ \ket{\omega_{10}}\ \ket{\omega_{01}}\ \ket{\omega_{00}}\\
    &\hspace{5ex}\ket{\omega_{11}^{-1}}\ \ket{\omega_{10}^{-1}}\ \ket{\omega_{01}^{-1}}\ \ket{\omega_{00}^{-1}}]\cdot \ket{\widetilde{\Phi}},
\end{align}
where $\ket{\widetilde{\Phi}}$ is denoted as
\begin{align}
    \ket{\widetilde{\Phi}} = \fraction{1}{\sqrt{2}}\twovec{\ket{\widetilde{\varphi}_1}\otimes\ket{\widetilde{\varphi}_2}}{\ket{\widetilde{\varphi}_1}\otimes\ket{\widetilde{\varphi}_2}}
\end{align}
with
\begin{equation}
    \ket{\widetilde{\varphi}_i} = \twovec{\sin \theta_i}{\cos \theta_i}\ (i\in \{1,\,2\}).
\end{equation}
We discuss the operation of $U$ with the basis $\{\ket{\omega_{ij}},\,\ket{\omega_{ij}^{-1}}\,|\,(i,\,j)\in \{0,\,1\}^2\}$.
The representative matrix of operators $U_0$ and $R_f$ for the basis $\{\ket{\omega_{ij}},\,\ket{\omega_{ij}^{-1}}\,|\,(i,\,j)\in \{0,\,1\}^2\}$ are obtained by $\widetilde{U}_0$ and $\widetilde{R}_f$ given as follows, respectively:
\begin{align}
    \widetilde{U}_0 &= \twobytwo{O_{4,4}}{I_2\otimes \widetilde{D}_{\varphi_2}}{\widetilde{D}_{\varphi_1}\otimes I_2}{O_{4,4}},\\
%    \widetilde{U}_0 &= \twobytwo{O_{4,4}}{I_2\otimes \widetilde{D}_{\varphi_2}}{\widetilde{D}_{\varphi_2}\otimes I_2}{O_{4,4}},\\
    \widetilde{R}_f &= \twobytwo{\widetilde{R_f}_\rP\otimes I_2}{O_{4,4}}{O_{4,4}}{I_2\otimes \widetilde{R_f}_\rP},
\end{align}
where $I_2$ is the $2\times 2$ identity matrix, $O_{4,4}$ is the $4\times 4$ zero matrix, and $\widetilde{D}_{\varphi_i}$ ($i \in \{0,\,1\}$) is a matrix defined as follows:
\begin{equation}
    \widetilde{D}_{\varphi_i} = 2\ketbra{\widetilde{\varphi}_i}{\widetilde{\varphi_i}} -I_2 = \twobytwo[r]{-\cos 2\theta_i}{\sin 2\theta_i}{\sin 2\theta_i}{\cos 2\theta_i}.
\end{equation}
Therefore, the representative matrix of the whole operation $\widetilde{U}$ is calculated as 
\begin{equation}
    \widetilde{U} = \widetilde{U}_0\widetilde{R_f} = \twobytwo{O_{4,4}}{I_2\otimes \widetilde{D}_{\varphi_2}\widetilde{R_f}_\rP}{\widetilde{D}_{\varphi_1}\widetilde{R_f}_\rP\otimes I_2}{O_{4,4}}.
\end{equation}
By the definition of representation matrices, $t$ steps of time evolution on the complete bipartite graph with the operator $U = U_0R_f$ can be calculated as
\begin{equation}
    \begin{split}
        U^t\ket{\Phi} &= [\ket{\omega_{11}}\ \ket{\omega_{10}}\ \ket{\omega_{01}}\ \ket{\omega_{00}}\\
    &\hspace{4ex}\ket{\omega_{11}^{-1}}\ \ket{\omega_{10}^{-1}}\ \ket{\omega_{01}^{-1}}\ \ket{\omega_{00}^{-1}}]\cdot \widetilde{U}^t\ket{\widetilde{\Phi}}.
    \end{split}
\end{equation}
Here, for a positive integer $s$, it holds that
\begin{align}
    &
    \begin{split}
        &\widetilde{U}^{2s-1}\ket{\widetilde{\Phi}} \\
        &= \fraction{1}{\sqrt{2}}\twovec{\twovec{\sin((2s-1)\theta_1)}{\cos((2s-1)\theta_1)}\otimes \twovec{\sin((2s+1)\theta_2)}{\cos((2s+1)\theta_2)}}{\twovec{\sin((2s+1)\theta_1)}{\cos((2s+1)\theta_1)}\otimes \twovec{\sin((2s-1)\theta_2)}{\cos((2s-1)\theta_2)}\vrule height18pt width0pt depth0pt},
%        &= \fraction{1}{\sqrt{2}}\twovec{\twovec{\sin((2s-1)\theta_1)}{\cos((2s-1)\theta_1)}\otimes \twovec{\sin((2s+1)\theta_1)}{\cos((2s+1)\theta_1)}}{\twovec{\sin((2s+1)\theta_1)}{\cos((2s+1)\theta_1)}\otimes \twovec{\sin((2s-1)\theta_1)}{\cos((2s-1)\theta_1)}\vrule height18pt width0pt depth0pt},
    \end{split}\\
    &
    \begin{split}
        &\widetilde{U}^{2s}\ket{\widetilde{\Phi}} \\
        &= \fraction{1}{\sqrt{2}}\twovec{\twovec{\sin((2s+1)\theta_1)}{\cos((2s+1)\theta_1)}\otimes \twovec{\sin((2s+1)\theta_2)}{\cos((2s+1)\theta_2)}}{\twovec{\sin((2s+1)\theta_1)}{\cos((2s+1)\theta_1)}\otimes \twovec{\sin((2s+1)\theta_2)}{\cos((2s+1)\theta_2)}\vrule height18pt width0pt depth0pt},
%        &= \fraction{1}{\sqrt{2}}\twovec{\twovec{\sin((2s+1)\theta_1)}{\cos((2s+1)\theta_1)}\otimes \twovec{\sin((2s+1)\theta_1)}{\cos((2s+1)\theta_1)}}{\twovec{\sin((2s+1)\theta_1)}{\cos((2s+1)\theta_1)}\otimes \twovec{\sin((2s+1)\theta_1)}{\cos((2s+1)\theta_1)}\vrule height18pt width0pt depth0pt},
    \end{split}
\end{align}
That is, the recommendation probability $P_{2s}(w)$ of arm $w$ after $2s$ steps of time evolution is calculated as follows: if $w\in V_2$,
\begin{widetext}
\begin{equation}\begin{split}
    P_{2s}(w) &= \sum_{\sigma'\in\Sigma}\sum_{\substack{v\in V_1\\ \sigma\in\Sigma}}|\braket{v,\,\sigma;\,w,\,\sigma'|U^{2s}|\Phi}|^2\\
    &= \fraction{1}{2}\sin^2((2s+1)\theta_1)\sin^2((2s+1)\theta_2)\sum_{\sigma'\in\Sigma_w^*}\sum_{\substack{v\in V_1\\ \sigma\in\Sigma_v^*}}|\braket{v,\,\sigma;\,w,\,\sigma'|\omega_{11}}|^2\\
    &\quad +\fraction{1}{2}\sin^2((2s+1)\theta_1)\cos^2((2s+1)\theta_2)\sum_{\sigma'\in\Sigma_w^*}\sum_{\substack{v\in V_1\\ \sigma\not\in\Sigma_v^*}}|\braket{v,\,\sigma;\,w,\,\sigma'|\omega_{10}}|^2\\
    &\quad +\fraction{1}{2}\cos^2((2s+1)\theta_1)\sin^2((2s+1)\theta_2)\sum_{\sigma'\not\in\Sigma_w^*}\sum_{\substack{v\in V_1\\ \sigma\in\Sigma_v^*}}|\braket{v,\,\sigma;\,w,\,\sigma'|\omega_{01}}|^2\\
    &\quad +\fraction{1}{2}\cos^2((2s+1)\theta_1)\cos^2((2s+1)\theta_2)\sum_{\sigma'\not\in\Sigma_w^*}\sum_{\substack{v\in V_1\\ \sigma\not\in\Sigma_v^*}}|\braket{v,\,\sigma;\,w,\,\sigma'|\omega_{00}}|^2\\
    &= \fraction{1}{2|V_2|}\left(1 +\fraction{(q_w-\overline{q}_2)(\sin^2((2s+1)\theta_2)-\overline{q}_2)}{\overline{q}_2(1-\overline{q}_2)}\right).
\end{split}\end{equation}
\end{widetext}
Equally, if $w\in V_1$ it holds that 
\begin{equation}
    P_{2s}(w) = \fraction{1}{2|V_1|}\left(1 +\fraction{(q_w-\overline{q}_1)(\sin^2((2s+1)\theta_1)-\overline{q}_1)}{\overline{q}_1(1-\overline{q}_1)}\right).
\end{equation}
Therefore, the same discussion as that in Appendix~\ref{app:complete} leads to the desired result.

\bibliography{tex_elements/tybib}
%\bibliography{../../../97_tex_elements/tybib}

\end{document}